\documentclass[prd,aps,preprint,amsmath,nofootinbib,amssymb,eqsecnum,showkeys,tightenlines]{revtex4-1}
\usepackage{slashed}
\usepackage{epsfig,latexsym,cancel,amssymb,amsmath,verbatim,mathrsfs}
\usepackage{color}
\usepackage{graphicx}
\usepackage{multirow}
\usepackage{amssymb}
\usepackage{float}%
\usepackage{hyperref}

\newcommand{\be}{\begin{equation}}
\newcommand{\ee}{\end{equation}}
\newcommand{\bea}{\begin{eqnarray}}
\newcommand{\eea}{\end{eqnarray}}
\newcommand{\ba}{\begin{array}}
\newcommand{\ea}{\end{array}}

\usepackage{tikz}
\usetikzlibrary{arrows,shapes}
\usetikzlibrary{trees}
\usetikzlibrary{matrix,arrows} 				
\usetikzlibrary{positioning}				
\usetikzlibrary{calc,through}				
\usetikzlibrary{decorations.pathreplacing}  
\usepackage{pgffor}							

\usetikzlibrary{decorations.pathmorphing}	
\usetikzlibrary{decorations.markings}
\tikzset{
	 >=stealth', 
    vector/.style={decorate, decoration={snake}, draw},
	provector/.style={decorate, decoration={snake,amplitude=2.5pt}, draw},
	antivector/.style={decorate, decoration={snake,amplitude=-2.5pt}, draw},
    fermion/.style={draw=black, postaction={decorate},
        decoration={markings,mark=at position .55 with {\arrow[draw=black]{>}}}},
    fermionbar/.style={draw=black, postaction={decorate},
        decoration={markings,mark=at position .55 with {\arrow[draw=black]{<}}}},
    fermionnoarrow/.style={draw=black},
    gluon/.style={decorate, draw=black,
        decoration={coil,amplitude=4pt, segment length=5pt}},
    scalar/.style={dashed,draw=black, postaction={decorate},
        decoration={markings,mark=at position .55 with {\arrow[draw=black]{>}}}},
    scalarbar/.style={dashed,draw=black, postaction={decorate},
        decoration={markings,mark=at position .55 with {\arrow[draw=black]{<}}}},
    scalarnoarrow/.style={dashed,draw=black},
   electron/.style={draw=black, postaction={decorate},
      decoration={markings,mark=at position .55 with {\arrow[draw=black]{>}}}},
	bigvector/.style={decorate, decoration={snake,amplitude=4pt}, draw},
}
\tikzstyle{block} = [draw, rectangle, 
    minimum height=3em, minimum width=6em]

\begin{document}

\title{Matter Asymmetries in the $Z_N$ Dark matter -companion Models}

\author{Shao-Long Chen}
\email[E-mail: ]{chensl@ccnu.edu.cn}
\affiliation{Key Laboratory of Quark and Lepton Physics (MoE) and Institute of Particle Physics, Central China Normal
University, Wuhan 430079, China}

\author{Zhaofeng Kang}
\email[E-mail: ]{zhaofengkang@gmail.com}
\affiliation{School of physics, Huazhong University of Science and Technology, Wuhan 430074, China}
 
\author{Ze-Kun Liu}
\email[E-mail: ]{liuzekun@htu.edu.cn}
\affiliation{Institute of Particle and Nuclear Physics, Henan Normal University, Xinxiang 453007, China}

\author{Peng Zhang}
\email[E-mail: ]{zhang\_peng961013@163.com}
\affiliation{School of physics, Huazhong University of Science and Technology, Wuhan 430074, China}

\date{\today}

\begin{abstract}

A class of $Z_{N\geq 3}$-symmetric WIMP dark matter models that are characterized by the semi-annihilation into the companion of dark matter has been proposed in Ref.~\cite{Guo:2021rre}, providing a mechanism to evade the stringent direct detection constraint. In this work, we point out that such models naturally provide the three Sakharov elements necessary for dark matter asymmetry, and moreover this asymmetry can be transferred to the visible sector with a proper link to the leptonic or quark sector. In our minimal $Z_3$ example, the migration to the leptonic sector is via the asymmetric companion decay into neutrinos, and the lepton asymmetry can be further transferred to the quark sector. The CP violation parameter is restrained in this model. Thus, we explore the thermal motion effect of dark matter and find that it gives an enhancement to the CP violation parameter, which is studied for the first time. A preliminary numerical analysis based on the Boltzmann equations shows that both correct relic density of dark matter and baryon asymmetry can be accommodated.


\end{abstract}

\maketitle

\newpage

\newpage

\section{Introduction}

Over the past few decades, various evidences have emerged to support the existence of dark matter (DM)~\cite{Bertone:2004pz,Rosati:2004mw,Krauss:1987kx}, which strongly supports the existence of new physics (NP) beyond the standard model (BSM).  The most widely studied DM candidates are the so-called weakly interacting massive particles (WIMPs),  which could naturally provide the right amount of dark matter relic density, a coincidence known as the WIMP miracle. However, the typical WIMP DM candidates are being besieged by many direct detection experiments, and the surviving room is approaching the neutrino floor. It questions the WIMP paradigm. A class of new WIMP DM candidates based on the $Z_N$ symmetric models, where DM possesses some symmetric companion into which, along with the Higgs boson, DM can semi-annihilation has been proposed in Ref.~\cite{Guo:2021rre}. In such models, the weak interactions that generate the DM annihilation process do not give rise to DM-nucleon scattering, thus leading to a WIMP DM free of direct detection signals. 

Moreover, the Baryon Asymmetry of the Universe (BAU) also calls for NP~\cite{shaposhnikov1992standard,dolgov2005cp}. The desired NP that is able to generate baryon asymmetry should satisfy a set of necessary properties, referred to as the three Sakharov Criteria~\cite{Sakharov:1967dj}. They are violation of the baryon number symmetry, violation of the discrete symmetries C (charge conjugation) and CP (the composition of charge conjugation and parity), and a departure from thermal equilibrium.

In general, understanding the origins of the visible matter asymmetry and dark matter develops along two parallel lines in the BSM. However, dark matter in the WIMP paradigm suggests that the two lines may have an intersection, as it precisely provides the two missing conditions of the three Sakharov Criteria: the freeze-out of thermal WIMP DM occurs in the cooling process of the Universe, which naturally provides the necessary departure from thermal equilibrium; in addition, sizable CP violation can be readily accommodated in the dark sector. Furthermore, by generalizing the baryon or lepton number to the dark sector, it is possible to violate baryon or lepton number perturbatively during the DM freeze-out epoch, which leads to DM annihilating baryogenesis or leptogenesis~\cite{Fukugita:1986hr,Dolgov:1991fr,Borah:2022uos,liu1993reexamination}. In particular, the latter scheme potentially has a deep connection with the origin of the neutrino mass, which is the most convincing evidence for new physics BSM. Such kind of exploration has been well studied in literature~\cite{Gu:2009yx,Cui:2011ab,Racker:2014uga,Cui:2015eba,Baldes:2015lka,Allahverdi:2022zqr,Elor:2018twp,Gogoi:2023jzl,Borah:2018uci,Chu:2021qwk,Choi:2023kxo}, which is dubbed as WIMPy baryogenesis. 

In Ref.~\cite{Guo:2021rre}, we realized that the characteristic semi-annihilation of DM violates a DM-number and therefore may give rise to the asymmetry between the dark matter particle and its antiparticle (the similar idea had also been studied in Ref.~\cite{Ghosh:2020lma,Ghosh:2023pkv}), and then it can be transferred to the visible sector by some simple extension. As a subsequent study, this work attempts to explicitly realize matter asymmetry in the $Z_N$-symmetric DM-companion model. We find that the built-in blocks of the $Z_N$-symmetric DM-companion model setup, which originally is not designed for matter asymmetry, surprisingly fit in the necessary elements for WIMPy baryogenesis.

As a concrete realization, we consider the $Z_3$ symmetric DM-companion benchmark model, where the Dirac fermion $\Psi$ is the DM candidate and the scalar $S$ is its companion, transforming as $\Psi/S\to e^{ik2\pi/3} \Psi/S$ under $Z_3$. Furthermore, we introduce a scalar doublet $\eta$ to link the dark sector and the leptonic sector; $\eta$ mixes with $S$, with mass eigenstates $S_{1,2}$. Such a simple setup is sufficient to realize the DM annihilating leptogenesis: the companion $S_1$ accumulates asymmetry from the asymmetric annihilation of DM until DM freeze-out and this asymmetry leaks into the leptonic sector through $S_1$ slow decay, free of washout effect from inverse decay. We make detailed numerical analysis by building our python code for the Boltzmann equations (BEs). To find that, for the heavy DM of several TeV, whose correct relic density relies on the annihilation with resonant enhancement from $S_2$, leptonic asymmetry can be further processed into baryon asymmetry via the active electroweak sphaleron process.

It is worth stressing that, in the minimal model, if one considers the CP violation originates from the interference between tree-level DM annihilation and mediators' self-energy correction, the CP violation will be suppressed when the running particles of the loop diagram are the DM particles themselves. Therefore, to obtain a nonvanishing CP-violation parameter, we need to take into account the thermal motion of the initial DM particles. We detailed this for the first time.


The paper is organized as follows: We present the mechanism and models in Section 2, clarifying the roles of DM particles and SM particles, respectively. In Section 3, we discuss the asymmetry generation in the DM annihilation process. Section 4 contains the numerical solution of the Boltzmann equations. Conclusions and discussions are in the last section. We add the details of calculations in the appendix, which may involve some subtleties.

\section{DM-companion models naturally seed matter asymmetry}

\subsection{Dark matter-companion equipped with a leptonic portal}
\subsubsection{Sakharov conditions for dark matter asymmetry generation}

In order to evade the constraints from DM direct detection, in Ref.~\cite{Guo:2021rre}, we have proposed a class of generalized Higgs-portal dark matter models that are based on the discrete DM protecting symmetry $Z_N$ with $N>2$. The key feature of this type of model is that dark matter has a $Z_N$ companion, a scalar field $S$ which has Higgs portal coupling and services as a bridge between SM and dark matter. In the limit of vanishing DM-Higgs-portal coupling, it enables dark matter to obtain the correct relic density by semi-annihilating into the companion plus Higgs boson, thus maintaining the role of WIMP dark matter candidate. The benchmark model is a $Z_3$ symmetric model for a fermionic DM $\Psi_R$ and its companion $S$, with the Lagrangian as
\begin{align}\label{model:hidden}
    -{\cal L}_{Z_3}\supset & + M_\Psi\bar{\Psi}\Psi+ \lambda_{sh}|S|^2H^{\dagger}H+ m_S^2|S|^2\\
 \nonumber
 &+ \left(\frac{A_s}{3}S^3+\lambda_{L}\overline{\Psi^{C}}P_L\Psi{S}+\lambda_{R}\overline{\Psi^{C}}P_R\Psi{S}+\text{h.c.}\right),
 \end{align}
where both $\Psi$ and $S$ are singlets under the SM gauge group. Note that the chiral partner $\Psi_L$ is introduced simply to make $\Psi_{R}$ massive, since its Majorana mass term is forbidden by the $Z_3$ symmetry. These two Weyl fermions form a massive Dirac fermion $\Psi\equiv (\Psi_L, \Psi_R)^T$. In this mechanism, $\Psi$ does not have any channel to generate DM-nucleon scattering according to Ref.~\cite{Guo:2021rre}. By contrast, the scalar DM requires suppression on DM-Higgs portal coupling by hand. Hence, we merely focus on the fermionic dark matter models. For other values of $N$ in the $Z_N$ symmetry, the above structure can be accommodated via similar field content with proper charge assignment~\cite{Guo:2021rre}. 

The above models are designed to evade dark matter direct detection, but they present a bonus that is beyond the conventional WIMP dark matter scenario. Let us first notice that the fermionic dark matter must be a Dirac fermion, and thereby can be assigned a dark matter number $U(1)_D$.  Then, we can deduce that they naturally satisfy the three Sakharov conditions~\cite{Sakharov:1967dj} for dark matter asymmetry generation: 
\begin{itemize}
    \item Dark matter number violation. It is manifest in the characteristic semi-annihilation channel for dark matter, ``$\rm DM + DM \to companion + Higgs~boson$".  The essential feature is the separation of dark matter and anti dark matter annihilation, which may have different cross-sections, as long as the following condition holds. 
    \item CP violation. The CP violating phase is readily present in the above dark sector: there are four independent complex parameters $M_\Psi$, $A_s$ and $\lambda_{L,R}$, and three of their phases can be absorbed by three fields $\Psi_{L,R}$ and $S$, leaving one physical phase. 
    \item Out-of-equilibrium. WIMP-like dark matter freeze-out naturally establishes the out-of-equilibrium environment at early Universe, to break the wash-out of dark matter asymmetry by inverse processes. 
\end{itemize}

\subsubsection{Adding a leptonic portal}

To transfer dark matter asymmetry to the visible matter asymmetry, an extra portal is needed to link the dark sector with the lepton or quark sector~\footnote{As a matter of fact, in Ref.~\cite{Guo:2021rre} we have noted that in the minimal model specified by Eq.~(\ref{model:hidden}) both DM and the companion are stable, due to the accidental $Z_2$ symmetry, and to make the companion decay one needs such a new portal.}, which makes the dark matter number turn out to be the generalized baryon or lepton number. The most economical and natural choice is introducing a doublet scalar $\eta$ that enables the direct coupling between the dark sector and leptonic sector of the SM. The total field content is listed in Table.~\ref{fields}, and the Lagrangian involving the newly added particles is:
\begin{align}\label{model:link}
 -{\cal L}_{extra}\supset &+y_{i}\overline{L_{Li}}\tilde{\eta}\Psi_{R}+\lambda_{H\eta}H^{\dagger}H\eta^{\dagger}\eta+g\eta^{\dagger}HS+m_\eta^2\eta^{\dagger}\eta,
 \end{align}
with $\widetilde \eta=i\sigma_2 \eta^*$.  We have ``derived" the scotogenic model in Ref.~\cite{Ma:2006km}, which originally is proposed to realize radiative neutrino masses at one-loop level. In this work, we will not focus on neutrino mass generation. Hence, $y_{i}$ can be set to very small, irrelevant to our discussion except that it furnishes a decay channel for the companion which otherwise is stable: $S\to \Psi_R+\nu_L$. In fact, $y_{i}\sim {\cal O}(1)$ returns to the scenario previously studied by others, where dark matter directly semi-annihilate into lepton, producing lepton asymmetry~\cite{Borah:2018uci}. 
\begin{table}[htb]
 \begin{center}
 \begin{tabular}{|c| c| c| c|} 
 \hline
Fileds & $SU(2)$ & $U(1)_Y$ & $~\mathcal{Z}_3~$ \\ \hline
 $H$   & 2  & 1/2 & 1 \\ 
 \hline
 $L_L$ & 2  & -1/2 & 1 \\ 
 \hline
 $\Psi^i_{R,L}$ & 1  & 0 &$\omega$ \\ 
 \hline
 $\eta=\begin{pmatrix}
     \eta^{\dagger}\\
     \eta^0\\
     \end{pmatrix}$ & 2 & 1/2 & $\omega$ \\
 \hline
 $S$  & 1 &  0  & $\omega$ \\
 \hline
 \end{tabular}
\end{center}
\caption{Fields and their representations under the symmetries.}\label{fields}
\end{table}

We add remarks on the possible generalization of the baryon number to the dark sector. A simple link to the quark sector using a scalar quark $\tilde q$ as the first term in Eq.~(\ref{model:link}) does not work, since $\tilde q$ cannot mix with $S$ due to the $SU(3)_c$ symmetry and hence baryon number cannot be generalized to the dark sector via the baryonic link. Actually, this is impossible with field content within the SM, since it admits the lowest baron number violating operator $\frac{1}{\Lambda^2}qqq\ell$, which violates lepton number at the same time thus leading to proton decay. However, in the supersymmetry model, there are available renormalizable baryon number violating operators, allowing for operators such as $\frac{1}{\Lambda}XU^cD^cD^c$, from which $X=S+\theta^2\Psi_L$ gains baryon number; $U^C$ and $D^C$ are quark superfields. This kind of model is of interest, and actually it even has an advantage for baryogenesis, compared to the leptogensis model above. However, the latter is elegantly connected to the neutrino mass issue, and thus in this work, we focus on it and leave the baryonic portal model in future studies.

\subsection{Particle spectrum and interaction couplings}
Besides the SM Higgs, the model involves two scalar fields, which are singlet scalar $S$ and doublet scalar $\eta$. These two new scalar fields are introduced as DM mediators. Let us decompose the two scalar doublets as
\begin{equation}
H=\dbinom{H^+}{H^0}\,, \quad \eta=\dbinom{\eta^+}{\eta^0},
\end{equation}
which are distinguished by their $Z_3$ charges. These scalar fields have the following $Z_3$-invariant scalar potential 
\begin{align}
V(H,\eta,S) =&m_H^2 H^{\dagger}H + m_{\eta}^2\,\eta^{\dagger} \eta + (m'_S)^{2} |S|^2 + \lambda_{H} \left(H^{\dagger}H\right)^2 + \lambda_{\eta} \left(\eta^{\dagger}\eta \right)^2 + \lambda_{S}\, |S|^4  \nonumber \\
&+ \lambda_{5}\, H^{\dagger} H \eta^{\dagger}\eta + \lambda_{4}\, H^{\dagger}H |S|^2 + \lambda_{\eta S}\,\eta^{\dagger} \eta |S|^2 \nonumber \\
&+ \left(f_1 S^3 + f_2\ \eta^{\dagger} HS + \text{h.c.}\right).
\end{align}
The mass terms of the neutral complex scalar fields $\eta^0$ and $S$ read
\begin{equation}
\mathcal{L}_{\text{mass}}=\left((\eta^0)^* , S^* \right)\begin{pmatrix}
m_{\eta}^2+\frac{1}{2}\lambda_5v^2 &  f_2 v/\sqrt2 \\
 f^*_2 v/\sqrt2 & (m'_S)^{2}+\frac{1}{2}\lambda_4v^2
\end{pmatrix}
\dbinom{\eta^0 }{S } \,.
\end{equation}
It is more convenient to define the elements as $m_{\eta^\pm}^2\equiv m_{\eta}^2+\frac{1}{2}\lambda_5v^2$ which is the charged boson mass squared, 
$m_S^2\equiv (m'_S)^{2}+\frac{1}{2}\lambda_4v^2$ and the mixing parameter $m_{\eta S}^2\equiv  f_2 v/\sqrt2$ which is taken to be real via rephasing $S$. 
The two physical fields in the mass basis are related to $\eta^0$ and $S$ via
\begin{equation}
\dbinom{S_1}{S_2}=\begin{pmatrix}
\cos\alpha  & \sin\alpha \\
-\sin\alpha & \cos\alpha
\end{pmatrix}
\dbinom{\eta^0}{S} ,
\end{equation}
where the mixing angle $\alpha$ is defined as
\begin{align}
      \sin\alpha =\frac{-m_S^2+m_{\eta^\pm}^2+\sqrt{(m_{\eta^\pm}^2-m_S^2)^2+4m_{\eta S}^4}}{8m_{\eta S}^4+2\lambda_{sh}^2v^4+2\lambda_{\eta}v^4-2(m_{\eta^\pm}^2-m_S^2)\sqrt{(m_{\eta^\pm}^2-m_S^2)^2+4m_{\eta S}^4}}.
\end{align}
The two states respectively have mass squared, 
\begin{align}\label{masses}
m_{S_{1,2}}^2\equiv\frac{1}{2}\left(m_{\eta^\pm}^2+m_S^2\mp\sqrt{(m_{\eta^\pm}^2-m_S^2)^2+4m_{\eta S}^4}\right),
\end{align}
To guarantee the access of companion channel for DM annihilation, the DM mass $m_\Psi$ should lie in the interval
\begin{align}\label{window}
    (m_{S_{1,2}}+m_h)/2<M_\Psi<m_{S_{1,2}},
\end{align}
with $m_{S_1}$ the lighter companion mass. In the heavy dark matter region, it gives the bound of the companion and dark matter mass ratio: $1<m_{S_1}/M_\Psi<2$. The charged Higgs boson $\eta^\pm$ is irrelevant here, assumed to be sufficiently heavy to satisfy the LHC bound.

Next, we collect the relevant interactions written in the mass basis. Consider a more general (or maybe a more realistic setup in the context of neutrino mass genesis at loop level) setup incorporating flavor indices for $\Psi_{L,R}$
\begin{equation}\label{Yukawa}
-\mathcal{L}_Y \supset  \lambda_{Rij}\,\overline{\Psi^{C}_{Ri}} \Psi_{Rj} S + \lambda_{Lij}\,\overline{\Psi^{C}_{Li}} \Psi_{Lj} S + y_{ij}\, \overline{L_{Li}} \tilde{\eta} \Psi_{Rj} + \text{h.c.} .
\end{equation}
In the mass basis, the above Yukawa couplings are written as 
\begin{align}\label{relevantL}
    -{\cal L}_Y\supset &(+\lambda_{Ri}s_\alpha \overline{N^C_{Ri}}N_{Ri}S_1+\lambda_{Ri}{c_\alpha}\overline{N^C_{Ri}}N_{Ri}S_2+\lambda_{Li}s_\alpha \overline{N^C_{Li}}N_{Li}S_1+\lambda_{Li}{c_\alpha}\overline{N^C_{Li}}N_{Li}S_2\\
    \nonumber
    &+y_{ij}{c_\alpha} \overline{\nu_{Li}}\Psi_{Rj}S_1^*-y_{ij}s_\alpha\overline{\nu_{Li}}\Psi_{Rj}S_2^*)+\text{h.c.},
\end{align}
with $s_\alpha\equiv\sin\alpha$, $c_\alpha\equiv\cos\alpha$ and $N_i$ are the representation of $\Psi_i$ in the mass basis. In addition, the companion fields have the following couplings to the SM Higgs boson and $Z$ boson, 
\begin{align}\label{relevantL:hZ}
    -{\cal L}_{EW}&\supset  v(\lambda_{sh}s^2_\alpha+\lambda_{\eta}c^2_\alpha+\sqrt{2}\frac{f_2}{v}s_\alpha{c_\alpha})h|S_1|^2+v(\lambda_{sh}c^2_\alpha+\lambda_{\eta}s^2_\alpha-\sqrt{2}\frac{f_2}{v}s_\alpha{c_\alpha})h|S_2|^2\\
     \nonumber
     &+v(\lambda_{sh}   s_\alpha{c_\alpha}-\lambda_{\eta} s_\alpha{c_\alpha}-\frac{\sqrt{2}}{2}\frac{f_2}{v}s^2_\alpha+\frac{\sqrt{2}}{2}\frac{f_2}{v}c^2_\alpha)S_1 S_2^*h\\
     \nonumber
     &+ig\frac{T^3-\sin{\theta_{\omega}Q}}{\cos{\theta_{\omega}}}Z^{\mu}(s^2_\alpha S_1\partial_{\mu}S_1^*+c^2_\alpha S_2\partial_{\mu}S_2^*+s_\alpha {c_\alpha} S_1\partial_{\mu}S_2^*+s_\alpha {c_\alpha} S_2\partial_{\mu}S_1^*)+\text{h.c.},
\end{align}
where the couplings to $Z$ boson come from the doublet companion $\eta$.

Our main concern is $N_1$, the lightest one that services as the DM candidate. For the heavier states, such as  $N_{2}$, they are almost irrelevant to our discussions, and thus their couplings to $S_{1,2}$ can be turned off. In Eq.~(\ref{Yukawa}), the $\lambda_{L}$- and $\lambda_R$-type of Yukawa couplings play similar roles, and then, for simplicity, we consider very small $\lambda_L$ scenarios. Now, we arrive at the minimal model and collect all the relevant couplings below
\begin{align}\label{EFT:final}
    -{\cal L}_{mini}\supset &(+\lambda_{R1}\overline{N_{1}^C}N_{1R}S_1+\lambda_{R2}\overline{N_{1}^C}N_{1R}S_2
       +y_{1}\overline{\nu}_{L1}N_{1R}S_1^*-y_{2}\overline{\nu}_{L1}N_{1R}S_2^*)+\text{h.c.}\\
     \nonumber  &+\lambda_{sh1}vh|S_1|^2+\lambda_{sh2}vh|S_2|^2+\left(\lambda_{sh12}v~hS_1 S_2^*+\text{h.c.}\right)\\
     \nonumber
     &+iZ^{\mu}(\beta_1S_1\partial_{\mu}S_1^*+\beta_2S_2\partial_{\mu}S_2^*+\beta_3S_1\partial_{\mu}S_2^*+\beta_4S_2\partial_{\mu}S_1^*)+\text{h.c.},
\end{align}
where we have rewritten the coupling parameters as the combination of the Lagrangian parameters specified in Eq.~(\ref{relevantL}) and Eq.~(\ref{relevantL:hZ}), and one can readily find their concrete expressions there. Note that the dimensionless couplings $\lambda_{sh1}$, etc.,  can be significantly larger than 1 if $f_2\gg v$.




\section{Dark/visible matter asymmetry generation}

Let us briefly describe the dynamics of a visible lepton or baryon asymmetry genesis in the model specified by Eq.~(\ref{model:hidden}) and Eq.~(\ref{model:link}). First, at a very high temperature $T\gg M_\Psi$, the dark and visible sectors are in well thermal equilibrium and hence no asymmetry is effectively created. Then, as $T$ decreases to about $m_{N_1}/10$, the anti-annihilation of $N_{1}N_{1}$ is decoupled. Then dark matter asymmetry begins to be generated during the decoupling epoch of $N_{1}N_{1}$ annihilation into the companion $S_1^*$ and $h$. The produced asymmetry of $S_1$ is immediately transferred to the leptonic sector via decay channel $S_1\to N_1+\nu_L$. Finally, the lepton asymmetry may have a chance of being transferred to the quark sector, on the condition that the dark matter is sufficiently heavy such that its decoupling occurs before the decoupling of the electroweak sphaleron. Therefore, it is necessary to consider both lepton asymmetry and baryon asymmetry separately, depending on the mass of dark matter. On the contrary, in the mentioned baryonic portal model, light dark matter is also able to generate a sufficiently large baryon asymmetry. 

In this section, we will calculate the CP-violation parameter, which suffers from the complexity caused by necessary thermal averaging in our model. Then, we derive the Boltzmann equations to describe the evolution of dark and visible matter asymmetry and give the qualitative analysis according to the solution. 

\subsection{CP violation parameter at zero temperature} 
To describe the asymmetry generated in the $N_{1}-$pair annihilation processes, we define the CP-violation parameter as in the decay scenario~\cite{Buchmuller:2004nz,Nardi:2007jp}:
\begin{align}\label{CPV:amp}
     \epsilon&=\frac{|M|^2_{N_{1}N_{1}\rightarrow S_1^*h}-|M|^2_{\bar{N}_{1}\bar{N}_{1}\rightarrow S_1h}}{|M|^2_{N_{1}N_{1}\rightarrow S_1^*h}+|M|^2_{\bar{N}_{1}\bar{N}_{1}\rightarrow S_1h}},
 \end{align}
where $|M|^2_{N_{1}N_{1}\to S_1^*h}$ denotes the modular square of matrix element for the process dark matter-dark matter annihilation into the $S_1^*$ and $h$, while the modular square of the matrix element of its CP conjugate process is $|M|^2_{\bar{N}_{1}\bar{N}_{1}\rightarrow S_1h}$. The process to $S_2^*$ and $h$ can be similarly included, but in general it is subdominant or even closed due to the heaviness of $S_2$, and thus its contribution is negligible. For completeness, we include the contribution of $S_2$ in the Boltzmann Equations (BEs) below. In general, due to the limitation of CPT symmetry and unitarity, the asymmetry of the total annihilation rate of $N_1N_1$ vanishes~\cite{Davidson:2008bu}. It means $\sum_{\{i=1,2\}}|{M}|^2_{N_{1}N_{1}\rightarrow S_i^*h}=\sum_{\{i=1,2\}}{M}|^2_{\overline{N_{1}}\overline{N_{1}}\rightarrow S_i^*h}$. But because of the presence of the heavier final state $S_2h$, part of annihilation rate of asymmetric~($|{M}|^2_{N_{1}N_{1}\rightarrow S_1^*h}-|{M}|^2_{\overline{N_{1}}\overline{N_{1}}\rightarrow S_1h}$)  does not vanish~\cite{Blazek:2021olf}.
 
The desired non-vanishing CP violation parameter $\epsilon$ asks for the interference between the tree-level and one-loop level graphs. In our model, the tree and one-loop graphs for the $N_{1}N_{1}$ $\to$ $S_1^*$ $h$ are depicted in Fig.~\ref{tree-loop1} and Fig.~\ref{tree-loop8}, with the tree level annihilation mediated by $S_1$ and $S_2$, respectively. For the one-loop corrections, we only need to include the self-energy contribution, because there is no vertex correction by virtue of the $Z_3$ symmetry and charge conservation~\cite{Dasgupta:2019lha}. Moreover, the self-energy corrections that contribute to $\epsilon$ should mediate different propagators, otherwise the resulting imaginary part of the coefficient factor to be zero.

Then, the four interference leads to $\epsilon=\epsilon_{12}+\epsilon_{14}+\epsilon_{34}+\epsilon_{32}$, where the subscript numbers correspond to the order of the four graphs specified in the captions. We quote the results below, with the details of calculation cast in the Appendix.~\ref{CPV},
\begin{align}\label{CPVs}
    \epsilon_{12} &=-\frac{{\rm Im}[\lambda_{R1}^*\lambda_{sh1}\lambda_{R2}\lambda_{R1}\lambda_{R2}^*\lambda_{sh1}]}{4\pi(s-m_{S_1}^2)^2(s-m_{S_2}^2)m_{N_{1}}^2}{\cal J},~~~~~~
     \epsilon_{32} = -\frac{{\rm Im}[\lambda_{R2}^*\lambda_{sh12}\lambda_{R2}\lambda_{R1}\lambda_{R2}^*\lambda_{sh1}]}{4\pi(s-m_{S_2}^2)^2(s-m_{S_1}^2)m_{N_{1}}^2}{\cal J},\\
     \nonumber
     \epsilon_{14} &=  -\frac{{\rm Im}[\lambda_{R1}^*\lambda_{sh1}\lambda_{R1}\lambda_{R2}\lambda_{R1}^*\lambda_{sh12}]}{4\pi(s-m_{S_1}^2)^2(s-m_{S_2}^2)m_{N_{1}}^2}{\cal J},~~~~~~
     \epsilon_{34} = -\frac{{\rm Im}[\lambda_{R2}^*\lambda_{sh12}\lambda_{R1}\lambda_{R2}\lambda_{R1}^*\lambda_{sh12}]}{4\pi(s-m_{S_2}^2)^2(s-m_{S_1}^2)m_{N_{1}}^2}{\cal J},
\end{align}
with the couplings defined in Eq.~(\ref{EFT:final}). The common factor  ${\cal J}$ is defined as
\begin{align}\label{suppressJ}
   {\cal J}\equiv \frac{\sqrt{s(s-4m_{N1}^2)}(s-2m_{N1}^2)}{\frac{|\lambda_{R1}\lambda_{sh1}|^2}{(s-m_{S_1}^2)^2}+\frac{|\lambda_{R2}\lambda_{sh12}|^2}{(s-m_{S_2}^2)^2}+\frac{\lambda_{R1}\lambda_{sh1}\lambda_{R2}^*\lambda_{sh12}+\lambda_{R1}^*\lambda_{sh1}\lambda_{R2}\lambda_{sh12}}{(s-m_{S_1}^2)(s-m_{S_2}^2)}}.
\end{align}
We will analyze these CP violation parameters in section~\ref{CPV:analysis} and discuss how to get resonantly enhanced when $2m_{N_1}\approx m_{S_2}$. Later, we will explain that this enhancement is necessary to realize baryogenesis in our model. 

For the convenience of our subsequent discussion, we consider two scenarios in light of the main channel of dark matter annihilation:


\begin{itemize}
    \item $S_1$ provides the dominant annihilation channel for the relatively light dark matter ($\lesssim 1$ TeV), and this case corresponds to pure leptogenesis with no need for resonance. Setting $\lambda_{sh12}$ sufficiently small or $m_{S_2}$ sufficiently heavy allows us to neglect the role of $S_2$ in determining DM relic density.  \begin{figure}[htbp]
\includegraphics[width=0.39\textwidth]{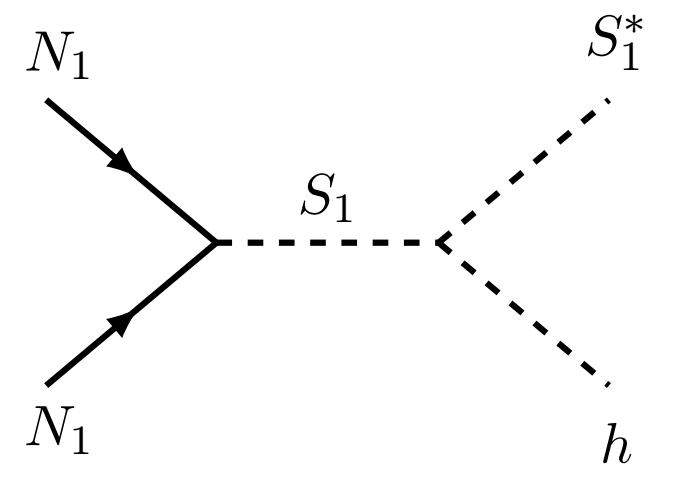}
\includegraphics[width=0.55\textwidth]{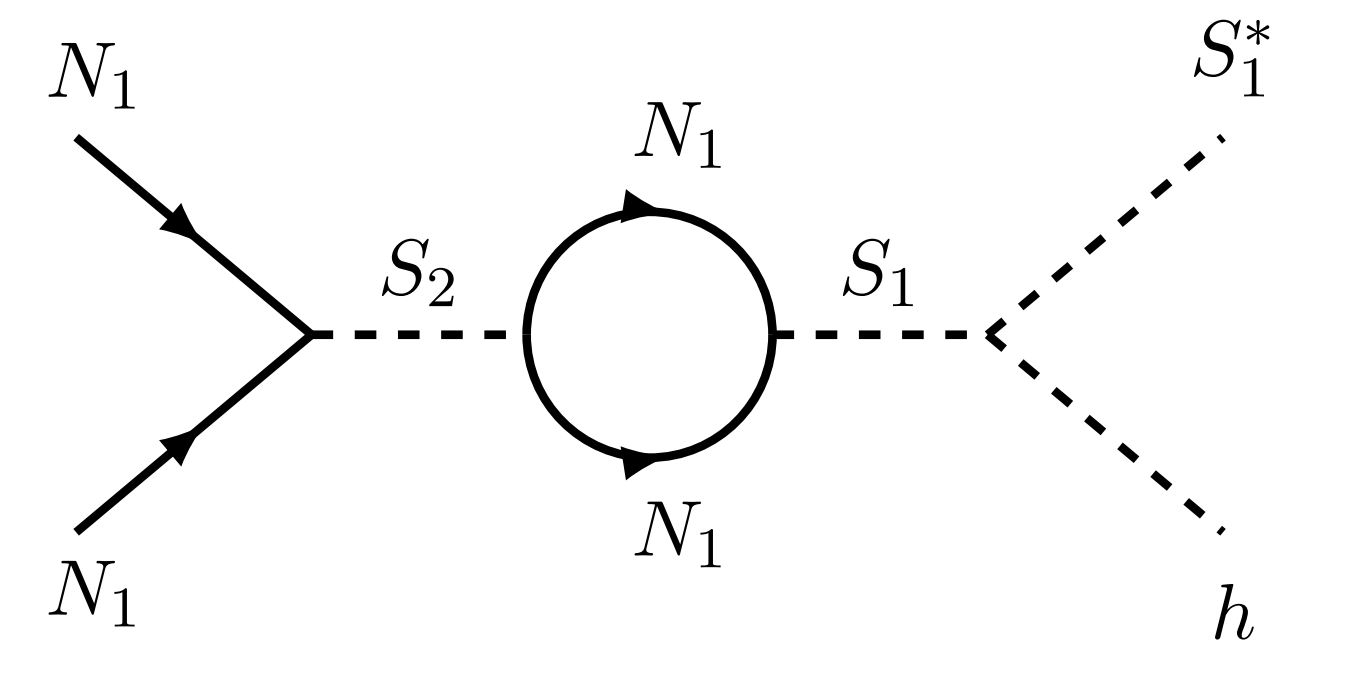}
\caption{The Feynman diagrams of annihilation $N_1N_1\to S_1^*h$ via the $S_1$ channel. Left: tree-level annihilation (labeled 1). Right: one-loop correction (labeled 2).}
\label{tree-loop1}
\end{figure}
    \item In contrast, $S_2$ provides the main annihilation channel for the multi-TeV scale heavy dark matter required for baryogenesis, and its mass needs to be adjusted to lie around the annihilation pole. We will find that $\lambda_{sh1}$ should still take a relatively large value to allow for resonant enhancement of the CP violation parameter, and hence the  $S_1$ channel may still give a subdominant contribution.
\begin{figure}[htbp]
\includegraphics[width=0.39\textwidth]{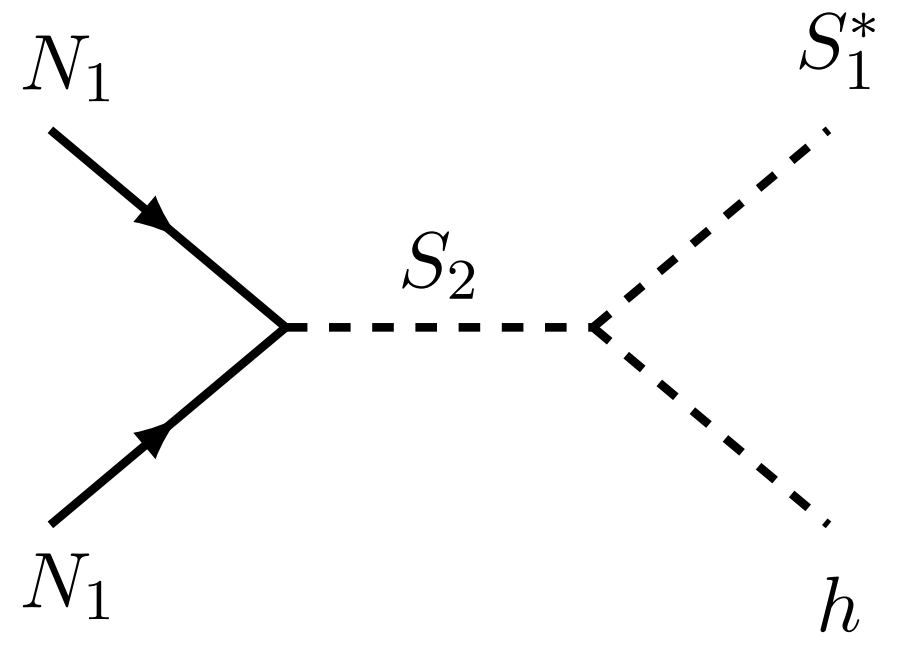}
\includegraphics[width=0.55\textwidth]{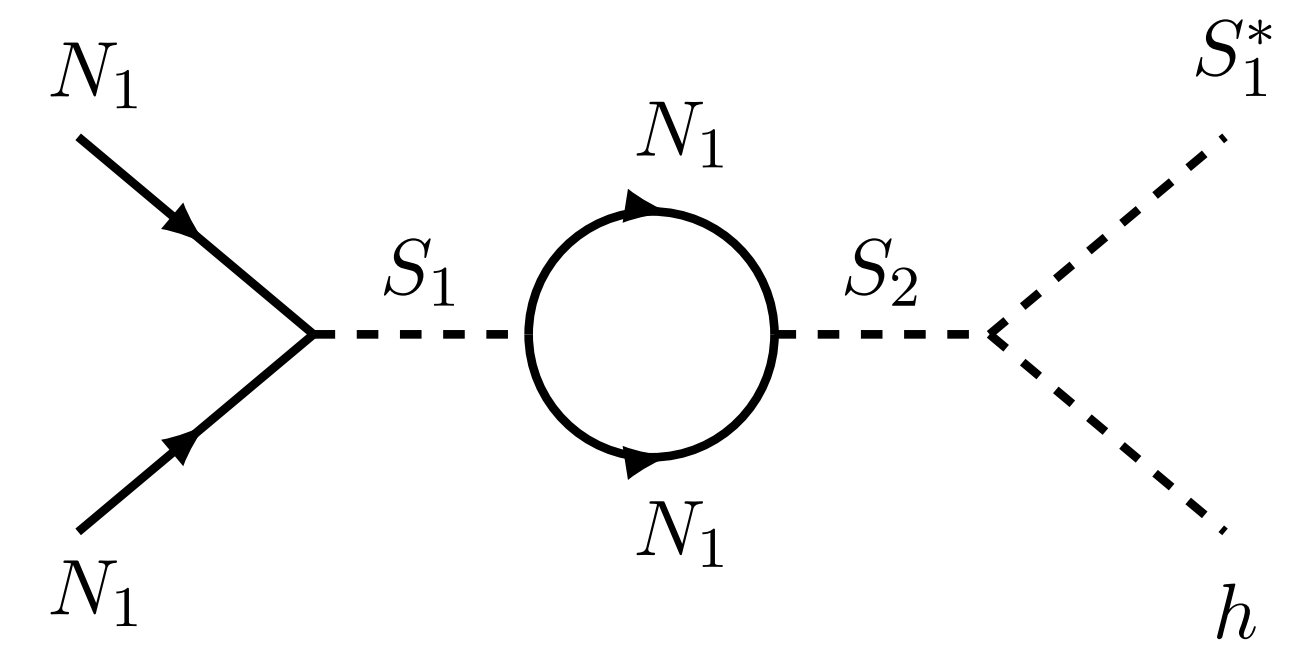}
\caption{The Feynman diagrams of annihilation $N_1N_1\to S_1^*h$ via the $S_2$ channel. Left: tree-level annihilation (labeled 3). Right: one-loop correction (labeled 4).}
\label{tree-loop8}
\end{figure}
\end{itemize}
We will consider non-resonant and resonant two scenarios respectively. A complete treatment may be necessary in some parameter space, but in general, it will not bring about significant differences in our simplifying discussions.

Let us move to the other channel with $h$ replaced by the $Z$ boson for the annihilation process $N_{1}N_{1}\rightarrow S_1^*Z$. The relevant Lagrangian reads
\begin{align}
    -{\cal L}\supset &+Z^\mu\beta_1(S_1\partial_\mu S_1 ^*)+\text{h.c.},
    \end{align}
which is given by the $D_{\mu}\eta^\dagger D^{\mu}\eta$. However, the result detailed in the appendix reveals that the process has no contribution to CP violation.

From Eq.~(\ref{suppressJ}) one can see that, we encounter a problem that $\epsilon$ vanishes as $s\to 4m_{N_{1}}^2$, which is the limit for two static dark matter particles annihilating. It is traced back to the fact that the one-loop process is via the loop where the same particles as the initial particle are running.  As a consequence, the strong CP phase, which originates from the on-shell of the loop particles, suffers a suppression. Then, in order to get a sizeable $\epsilon$, we turn to the thermal motion of dark matter during decoupling~\footnote{Beyond the minimal model Eq.~(\ref{EFT:final}), we have other ways out of this difficult. For instance, if $y_{ij}$ is sizable, one may instead consider the loop process is via SM neutrino and dark matter. Alternatively, we can replace the initial $N_1$ with the heavier families such as $N_2$. }. The resulting CP violation parameter for a pair of colliding dark matter particles with generic four-momentum $p_1$ and $p_2$ is denoted as $\epsilon(p_1,p_2)$. This is not a trivial and widely discussed topic, and it is more convenient to study it after introducing BEs.

\subsection{Boltzmann Equations and Numerical Result}
In order to quantitatively study the final abundance of asymmetric dark matter and the number of leptons or baryons, we need to turn to BEs and their numerical solutions. It is convenient to cast the BEs in terms of $Y_X\equiv n_X/s$, the number density per comoving volume of species $X$ with mass $m_X$; $s$ is the entropy density. We define $x=m_{X}/T$ and use $Y_X^{eq}$ to indicate the abundance of  $X$ is in the thermal equilibrium.


\subsubsection{Derivation of BEs}



To facilitate the derivation of BEs for complex particle systems, we adopt the notation in Ref.~\cite{Nardi:2007jp}, which defines the ratio $y_X\equiv{Y_X}/{Y_X^{eq}}$ to describe deviation from equilibrium for the particle species $X$. The collision term involves $\gamma^A_B\equiv\gamma(A\rightarrow B)$, the interaction density for the process $A\to B$, which is defined as~\cite{Davidson:2008bu,Buchmuller:2004nz}
\begin{align}\label{reactionrate}
   \gamma^{a b \ldots}_{i j \ldots}&\equiv \gamma(a+b+\ldots\rightarrow i+j+\ldots)\\
   \nonumber
   &=\int d\Pi_{a}f^{eq}_{a}d\Pi_{b}f^{eq}_{b}\ldots|\mathcal{M}(a+b+\ldots\rightarrow i+j+\ldots)|^{2}\tilde{\delta}d\Pi_{i}d\Pi_{j},
\end{align}
where $d\Pi\equiv {d^3p}/{(2\pi)^3}{2p^0}$ is the invariant phase space and $\tilde \delta=(2\pi)^4\delta^4(P_i-P_f)$ denotes the momentum conservation. We have used the relationship $f/n=f^{eq}/n^{eq}$ to ensure that the distribution functions in the density $\gamma$ take the form of equilibrium one.

Then, the collision term is the summation of various following difference
\begin{align}\label{dif}
[A\leftrightarrow B]\equiv(\prod_{i=1}^n y_{A_i})\gamma^A_B-(\prod_{i=1}^m y_{B_i})\gamma^B_A,
\end{align}
where $A_i$ runs over the particle species in the initial state collectively denoted as $A$; similarly for $B_i$. Now, the reaction density can be written in a more illustrative manner, 
\begin{align}\label{rate}
\gamma^A_B\equiv  \prod_{i=1}^n n^{eq}_{A_i} \left \langle\sigma_{a+b+\ldots\rightarrow i+j+\ldots} v\right \rangle.
\end{align}
Note that the thermal average is understood as a cross section only for the case with the two-body initial state, while for the decay process, it is the thermally averaged decay rate $\langle \Gamma_{A\to B}  \rangle$. Anyway, the quantities in the right-handed of Eq.~(\ref{dif}) can be expressed as the following,
\begin{align}
(\prod_{i=1}^n y_{A_i})\gamma^A_B=s^{n}(\prod_{i=1}^n Y_{A_i})\left \langle\sigma_{a+b+\ldots\rightarrow i+j+\ldots} v\right \rangle.
\end{align}

With these definitions, we are able to derive the Boltzmann equations that describe the evolution of the number densities of DM, its companion and their antiparticles, namely $N_{1}$, $\bar{N}_{1}$, $S_i$ and $S_i^*$. Their evolution are controlled by the collision terms and the evolution equations are written as
\begin{align}\label{BEs:compact1}
    \dot{Y}_{S_1^*}=&[N_{1}N_{1}\leftrightarrow S_1^*h]+[\bar{}N_{1}\nu_1\leftrightarrow S_1^*]+[\bar{N}_{1}\nu_1\leftrightarrow S_1^*h]+[{\rm SM}~{\rm SM}\leftrightarrow S_1^*S_1],\\
    \dot{Y}_{S_1}=&[\bar{N}_{1}\bar{N}_{1}\leftrightarrow S_1h]+[N_{1}\overline{\nu_1}\leftrightarrow S_1]+[N_{1}\overline{\nu_1}\leftrightarrow S_1h] +[{\rm SM}~{\rm SM}\leftrightarrow S_1^*S_1],\\
    \dot{Y}_{S_2^*}=&[N_{1}N_{1}\leftrightarrow S_2^*h]+[\bar{N}_{1}\nu_1\leftrightarrow S_2^*]+[\bar{N}_{1}\nu_1\leftrightarrow S_2^*h] +[{\rm SM}~{\rm SM}\leftrightarrow S_2^*S_2],\\
    \dot{Y}_{S_2}=&[\bar{N}_{1}\bar{N}_{1}\leftrightarrow S_2h]+[N_{1}\overline{\nu_1}\leftrightarrow S_2]+[N_{1}\overline{\nu_1}\leftrightarrow S_2h] +[  {\rm SM}~{\rm SM}\leftrightarrow S_2^*S_2],\\
    \dot{Y}_{N_{1}}=&2[S_1^*h\leftrightarrow N_{1}N_{1}]+[S_1 \leftrightarrow N_{1}\overline{\nu_1}]+[S_1h \leftrightarrow N_{1}\overline{\nu_1}]+2[S_2^*h\leftrightarrow N_{1}N_{1}]\nonumber\\
    &+[S_2 \leftrightarrow N_{1}\overline{\nu_1}]+[S_2h \leftrightarrow N_{1}\overline{\nu_1}],\\
    \dot{Y}_{\bar{N}_{1}}=&2[S_1h \leftrightarrow \bar{N}_{1}\bar{N}_{1}]+[S_1^*\leftrightarrow \bar{N}_{1}\nu_1]+[S_1^*h\leftrightarrow  \bar{N}_{1}\nu_1]+2[S_2h \leftrightarrow \bar{N}_{1}\bar{N}_{1}] \nonumber \\
    &+[S_2^*\leftrightarrow \bar{N}_{1}\nu_1]+[S_2^*h\leftrightarrow  \bar{N}_{1}\nu_1].\label{BEs:compact2}
\end{align}
where we define $\dot{Y}$ as $\frac{s H(m)}{z} \frac{\mathrm{d}Y}{\mathrm{d}z}$, such as $\dot{Y}_{N_1}\equiv\frac{s H(m_{N_1})}{z} \frac{\mathrm{d}Y}{\mathrm{d}z}$ with $H(m_{N_1})$ the Hubble parameter at $T=m_{N_1}$. One needs to expand the compact terms further to recover the lengthy form of various thermal averages which are related to the following 
 \begin{equation}\label{crossCPV}
 \left \langle \sigma_{\overline{N_1}\overline{N_1}\rightarrow S_1h} v\right \rangle=\left\langle \sigma_{S_1^*h\rightarrow N_1 N_1} v\right \rangle
 \approx 
 (1-2\epsilon)\left \langle\sigma_{N_1N_1\rightarrow S_1^*h} v\right \rangle.
 \end{equation}
We put the details in Eq.~(\ref{BEs:expand1}-\ref{BEs:expand2}) in the appendix.  

The CP violation parameter $\epsilon$ present is the one in the static limit and therefore vanishes in our model as mentioned previously.  This means that the asymmetry source is absent and thus we are forced to include the thermal motion effect later.

\subsubsection{Dark matter freeze-out with only kinematic equilibrium  }

\begin{figure}[h]
\centering
\includegraphics[width=0.47\textwidth]{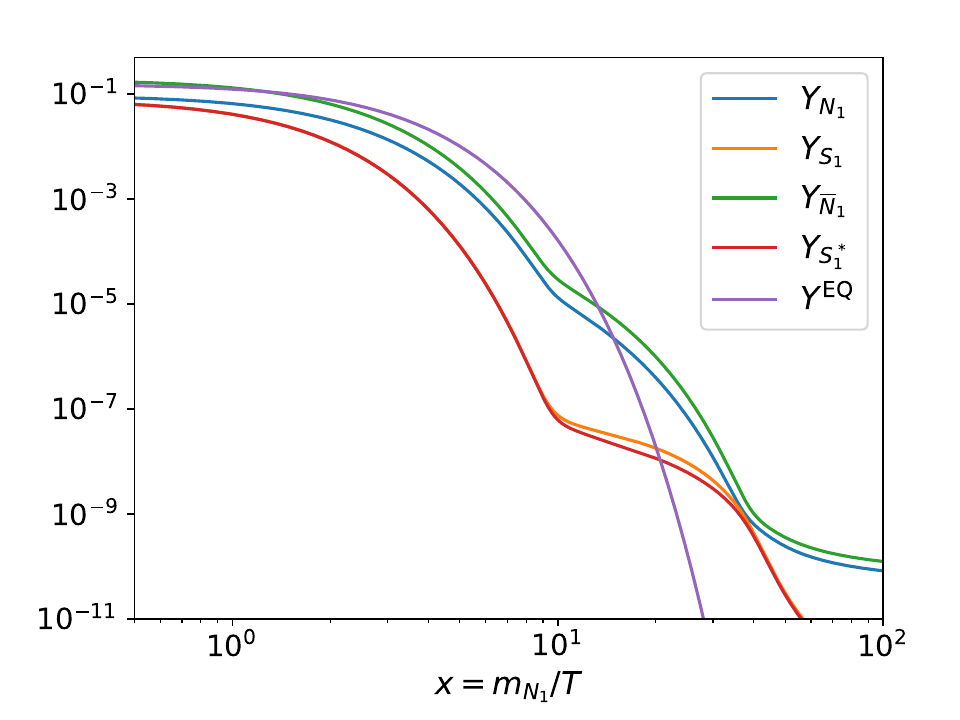}~
\includegraphics[width=0.47\textwidth]{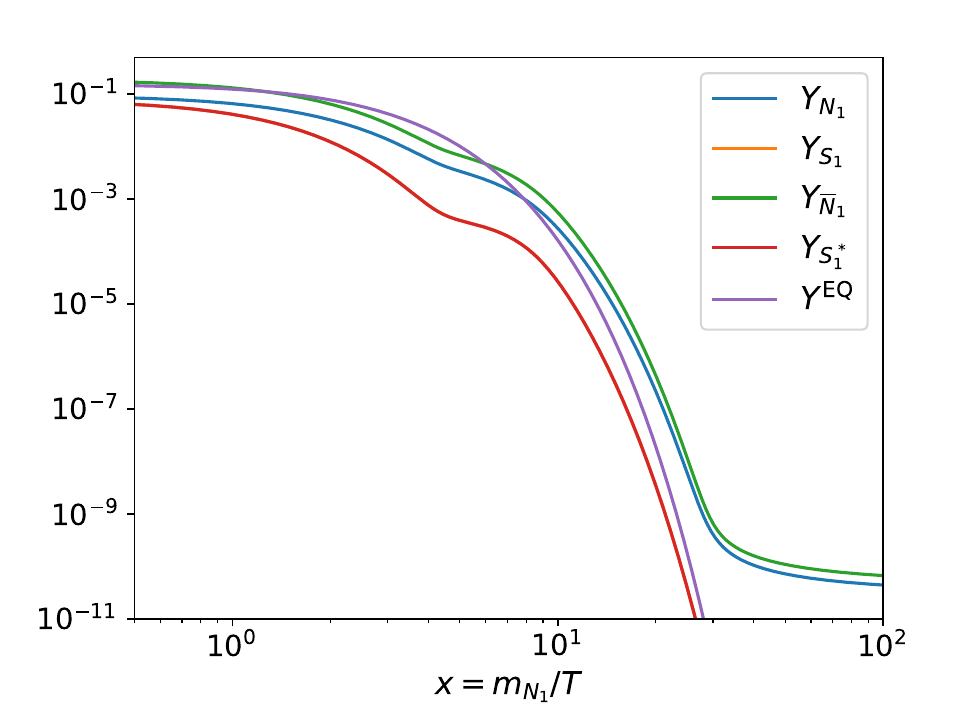}
\caption{Demonstration of the evolution of particle abundances. We fix dark matter mass $m_{N_1}=500$ GeV, CP-violation parameter $\epsilon=0.25$ and $\left\langle\sigma_{N_{1}N_{1}\rightarrow S_1^*h}v\right \rangle= 5\times 10^{-12}$ $\rm GeV^{-2}$ and $\left\langle\sigma_{S_{1}S_{1}^*\rightarrow hh}v\right \rangle =5\times 10^{-10}$ $\rm GeV^{-2}$. For other parameters, we set:  $m_{S_1}/m_{N_1}=1.6$ and $S_1$ decay width $\Gamma_D= 10^{-15}$ $\rm GeV$ (left panel), $m_{S_1}/m_{N_1}=1.9$ and $\Gamma_D=5\times 10^{-9}$ $\rm GeV$ (right panel). The change of $Y_{S_1}$ at $x=10$ for both $S_1$ and $N_1$ corresponds to the freeze-out of $S_1$. For $x>30$, $Y_{S_1}$ decreases sharply, it corresponds to the decay $S^*_1\to N_1\nu$(in the left). In the right plot, the kinks around $x=4$ correspond to the freeze-out of $S_1$. }
\label{evolution2}
\end{figure}

\begin{figure}[htbp]
\centering
\includegraphics[width=0.47\textwidth]{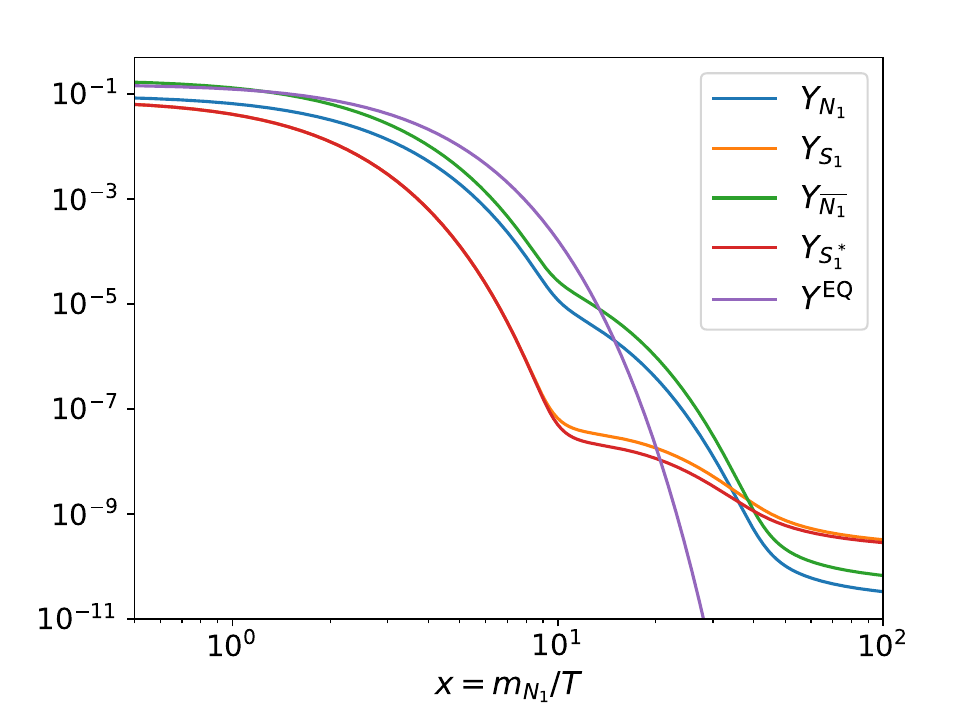}
\includegraphics[width=0.47\textwidth]{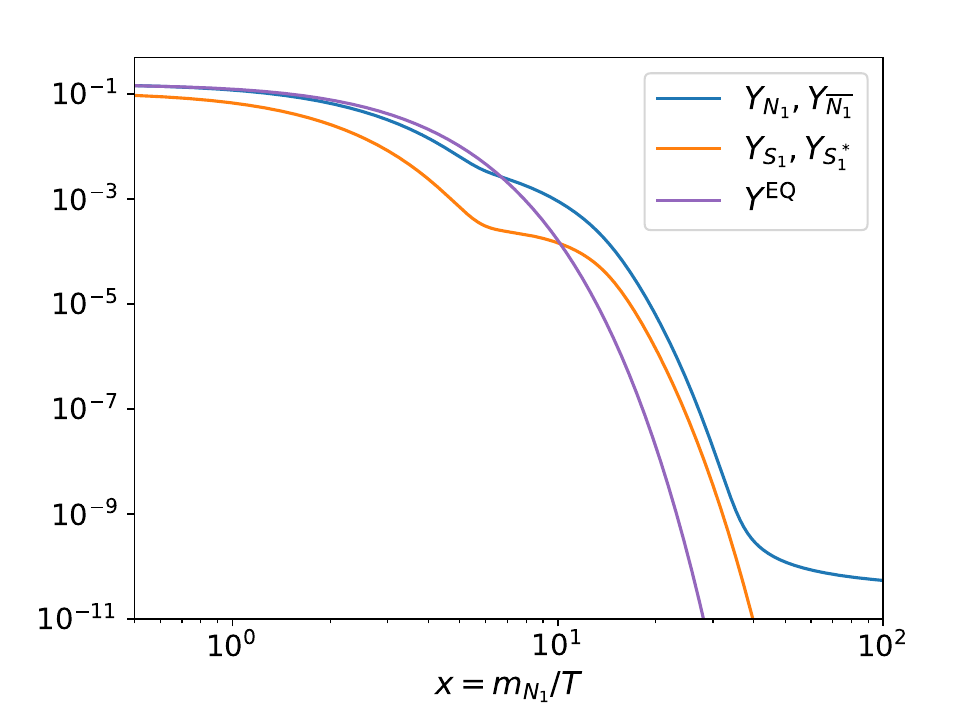}
\caption{Demonstration of the impacts of CP violation and/or $S_1$ decay width on the evolution of particle abundances. We set $\Gamma_D=0$, $\epsilon=0.25$ (left panel) and  $\Gamma_D=5\times 10^{-9}$GeV, $\epsilon=0$  (right panel) . $m_{S_1}/m_{N_1}=1.6$. The others are the same as in the previous figure. In the left plot, $Y_{S_1}$ decreases to 0 no longer owing to $\Gamma_D=0$. }
\label{evolution3}
\end{figure}

In this part, we give an illustrative discussion on the solution of Eq.~(\ref{BEs:compact1}-\ref{BEs:compact2}), showing how our mechanism of leptogenesis works. The schematic solution is plotted in Fig.~\ref{evolution2} and Fig.~\ref{evolution3}.
As the Universe cools down, $S_1$ ($S_2$ can be similarly included if necessary) decouples from the SM plasma first, as the annihilation rate $S_1S_1\to hh$, etc, drops below the Hubble expansion rate at $T_{f,S_1}\sim m_{S_1}/20\approx T'_{f,N_1}m_{S_1}/m_{N_1}$ with $T'_{f,N_1}\sim m_{N_1}/20$ (the usual estimation for dark matter freeze-out). Of course, dark matter also decouples from it. Hereafter, the two sectors lose chemical equilibrium, but they still share the same temperature due to the much later kinematic decoupling between them. Then, dark matter number density freezes out as usual inside the dark sector, when $T$ decreases to the value so that the semi-annihilation chain within the dark sector $N_1N_1\to S^*_1h\to N_1\nu+ h$ is frozen. It is estimated that this event happens at 
\begin{align}
    T_{f,N_1}\approx T_{f,S_1}-T'_{f,N_1}\sim (m_{S_1}-m_{N_1})/20\Rightarrow x_{f,N_1}\sim \frac{20}{m_{S_1}/m_{N_1}-1}.
\end{align}
In the heavy dark matter region $1<m_{S_1}/m_{N_1}<2$, and thus $x_{f,N_1}$ is more than 20. Therefore, the decoupling seems delayed. The above features that we observed are consistent with the schematic plots shown in Fig.~\ref{evolution2}, which take $m_{S_1}/m_{N_1}=1.6$ and 1.9 for the left and right panels, respectively, sharing other parameters except for the irrelevant decay width of $S_1$ (see the caption). This is not good news for realizing baryogenesis in our model which will be discussed later. For the precise solution of the decoupling temperature of dark matter, we can assume that decoupling occurs when $Y_{N_1}(x_f)$ drops to the same level as $Y_{N_1}(\infty)$.

In order to obtain the decoupling point, we give a graph of the slope of $Y_{N_1}(x)$. It can be seen that the freezeout of dark matter occurs when $\frac{m_{N1}}{T}$ is below 30.


\subsubsection{Leptogensis via the companion leaking: no wash-out limit}

Our mechanism shows a different feature than the usual direct annihilation leptogenesis mechanism: in our case, the leptonic asymmetry comes from the leak of heavier companion decay: the asymmetric decays between $S_1\to \bar\nu+N_1$ and $S_1^*\to \nu+\bar N_1$. To make the transfer effective, we should suppress the washout effect from inverse decay. Then, the relevant coupling $y_1$ in Eq.~(\ref{EFT:final}) which accounts for the $S_1$ decay is required to be sufficiently small, and thus the inverse decay is decoupled fairly early at $T_{ID}$, achieving the maximum conversion efficiency. Before proceeding to calculate matter asymmetry, we should study the precise condition for this.

Let us estimate $T_{ID}$ via the condition that the inverse decay rate drops below the Hubble expansion rate,
\begin{align}
  \Gamma_{I}(\bar\nu+N_1\to S_1)\approx e^{-\frac{m_{S_1}-m_{N_1}}{T}}\Gamma<H(T)=g_{*}^{1/2}\frac{T^{2}}{M_{P}}\left(\frac{4\pi^{3}}{45}\right)^{1/2},
\end{align}
where $\Gamma={y_1^2(m_{S_1}^2-m_{N_1}^2)^2}/{16\pi}{m_{S_1}^3}$ is the decay width of $S_1$ and $g_*$ is the relativistic degree of freedom in the SM around the weak scale. Then, we get the upper bound on $y_1$: 
\begin{align}
   y^2_1\lesssim4\pi g_{*}^{1/2}\left(\frac{4\pi^{3}}{45}\right)^{1/2} \frac{m_{S_1}}{M_{P}}\left(
   \frac{T_{ID}}{\Delta m}\right)^2e^{\Delta m/T_{ID}}\approx 4.6\times 10^{-11}\times\frac{m_{S_1}}{1\rm TeV}\left(\frac{T_{ID}/\Delta m}{0.1}\right)^2,
\end{align}
with $\Delta m=m_{S_1}-m_{N_1}$. For the even larger $y_1$, the washout effect from inverse decay becomes important and one should turn to numerical integration.

On the other hand, $y_1$ cannot be too small, otherwise the $S_1$ decay will happen too late to allow for successful baryogenesis; this requires its lifetime  $1/\Gamma <1/H(T=m_W)$, namely decay away around the weak scale. 
\begin{align}
   y^2_1\gtrsim  2.1\times 10^{-15}\times\frac{m_{S_1}}{1\rm TeV}\left(\frac{m_{W}/\Delta m}{0.1}\right)^2.
\end{align}
In summary, the coupling $y_1$ should lie within the region specified above for the validation of our subsequent analysis. 

\subsubsection{Leptogenesis via the companion leaking: counting the matter asymmetry}

Both the invisible and visible matter asymmetries are generated during the epoch of dark matter freeze-out. For ease of description, we define the point $A$ as the temperature point of washout freeze out and point $B$ as the temperature point of $N_{1}N_{1}$ annihilation freeze out. After washout freezes out, a net asymmetry of $S_1$ is present in the Universe. Then, the asymmetric decay of the companion, without wash-out as discussed previously, generates an asymmetry that can be accumulated in the visible sector. Because the process of CP-violation does not continue after $N_{1}N_{1}$ annihilation freeze out at the point $B$, the genesis of asymmetry loses its source in both sectors. After that, asymmetry in the invisible sector will completely immigrate to the visible sector.

From the above analysis, it is seen that we can estimate the total lepton asymmetry in this model, provided that we can figure out $Y_{S_1T}$ and $Y_{S_1^*T}$ describing the total number of $S_1$ and $S_1^{*}$ decay respectively. In our model, it receives two contributions, namely $Y_{ST}=Y_{S_1A}+\Delta Y_{S_1g}$, with $S_{1A}$ the initial value at $A$ which can be read directly from the figure. However, $\Delta Y_{S_1g}$, the historical production of abundance of $S_1$ from $A$ to $B$ cannot be read directly from the figure. Here, $\Delta Y_{S_1g}$ just refers to the net production from the pair annihilation $\bar{N}_{1}\bar{N}_{1}\to S_1 h$, whose inverse process tends to be suppressed when moving from point $A$ to $B$, so we specify to this net annihilation. In the following, we will relate $\Delta Y_{S_1g}$ to the change of dark matter abundance. 

On the evolution curve of a certain particle species $X$, the difference in $Y_X$ between two points reflects the accumulation of the difference between the production and destruction of $X$ during this historical period. For instance, for $S_1$ we have
\begin{align}\label{sd}
  Y_{S_1A}-Y_{S_1B}=\Delta Y_{S_1d}-\Delta Y_{S_1g},
\end{align}
with $\Delta Y_{S_1d}$ the destruction of $S_1$ by decay.  Note that the only way to produce $S_1$ is via $\bar{N}_{1}\bar{N}_{1}\to S_1 h$, so one simply has $2\Delta Y_{S_1g}=\Delta Y_{\bar{N}_1a}$ with $\Delta Y_{\bar{N}_1a}$ the abundance of $\bar{N}_{1}$ that has annihilated away during this process. But at the same time, it gains abundance from $S_1^*$ decay, and thus 
\begin{align}
   Y _{\bar{N}_1A}-Y_{\bar{N}_1B}=\Delta Y_{\bar{N}_1a}-\Delta Y_{S^*_1d}=2\Delta Y_{S_1g}-\Delta Y_{S^*_1d}=2\Delta Y_{S_1g}-(Y_{S^*_1A}-Y_{S^*_1B}+\Delta Y_{S^*_1g}).
\end{align}
With this relationship, we get 
\begin{align}
    2Y_{S_1T}&=2(Y_{S_1A}+\Delta Y_{S_1g})=2Y_{S_1A}-Y_{S_1B}^*+Y_{S_1A}^*+(Y_{\bar{N}_1A}-Y_{\bar{N}_1B})+\Delta Y_{S_1g}^*.
\end{align}
By utilizing $Y_{S_1^*T}=Y_{S^*_1A}+\Delta Y_{S_1^*g}$, the unknown intermediate quantity $\Delta Y_{S_1^*g}$ can be eliminated, leading to $2Y_{S_1T}-Y_{S_1^*T}=Y_{\bar{N}_1A}-Y_{\bar{N}_1B}+2Y_{S_1A}-Y_{S_1B}$. Eventually, we get the expression of neutrino asymmetry
\begin{align}
    (Y_{\nu}-Y_{\bar \nu})_A-(Y_{\nu}-Y_{\bar \nu})_B= Y_{S_1T}-Y_{S_1^*T}&\approx\frac{1}{3}[Y_{\bar{N}_1A}-Y_{\bar{N}_1B}-Y_{N_1A}+Y_{N_1B}],
\end{align}
where we have neglected the asymmetry change of $S_1$, which is found to be comparatively small. Note that the analysis is similar to the particle $S_2$.



\subsubsection{From lepton to baryon asymmetry}

In the previous subsections, we show how the lepton asymmetry is generated from the dark sector. It is important to investigate if the lepton asymmetry can be transferred to the baryon sector via the active electroweak sphaleron process. Owing to the electroweak phase transition,  the baryon asymmetry is derived from the $B-L$ asymmetry~\cite{Harvey:1990qw}. The baryon asymmetry is related
to the $B-L$ asymmetry by
\begin{align}
    \frac{n_{B}}{s}=\frac{24 + 4n_H}{66+13n_H}\frac{n_{B-L}}{s},
\end{align}
where $n_H$ is the number of Higgs doublets. The process converts the matching asymmetry $Y_{\Delta L}$ into a baryon asymmetry

\begin{align}
  Y_{\Delta B}  =\frac{28}{79}Y_{\Delta L}.
\end{align}
As the temperature decreases,  sphalerons continuously decouple. To ensure that the asymmetry of the visible sector can be preserved, we need the decoupling freeze to be later than the freeze-out of dark matter annihilation. Before sphaleron is fully decoupled, the production of baryon number is active. According to Ref.~\cite{Hong:2023zrf,DOnofrio:2014rug,kharzeev2020sphalerons}, the point of sphaleron freeze-out is noted by $T_{\text{FO}}\approx130$ GeV which is extrapolated from lattice simulation. As Fig.~\ref{evolution2} shows, the time of $N_{1}$ annihilation freezes out was earlier than the sphaleron freeze-out.

\subsection{Thermal motion effect on the CP violation parameter  }

\subsubsection{The leading thermally averaged CP violation annihilation parameter }

Now, we can provide a more accurate explanation of the suppressed $\epsilon$ problem that we have confronted before. In the BEs, we study the evolution of number densities and integrate over the momentum of the involved particles in the collision terms, see Eq.~(\ref{reactionrate}). There, we simply treat $\epsilon(p_1,p_2)$ as the constant in the static limit $\epsilon(s\to 4m_{N_{1}}^2)$, and consequently it can be factorized out from the integration as a numerical input, see Eq.~(\ref{CPV:BE}). However, once we try to go beyond that limit of $\epsilon(p_1,p_2)$ to take into account the thermal motion\cite{Covi:1997dr,Garbrecht:2013iga,Garbrecht:2013iga,giudice2004towards}, the exact treatment should include $\epsilon(p_1,p_2)$ in the integration, leading to the (Lorentz invariant) reaction density difference due to CP asymmetry
\begin{align}
   \delta\gamma^{a b}_{i j }&=\int d\Pi_{a}f^{eq}_{a}d\Pi_{b}f^{eq}_{b}\epsilon(a+b\rightarrow i+j)\int|\mathcal{M}_{tree}(a+b\rightarrow i+j)|^{2}\tilde{\delta}d\Pi_{i}d\Pi_{j}.
\end{align}
Formally, up to a numerical factor, it is approximated to be $n_a^{eq}n_b^{eq}\langle \sigma_{ab\to ij} v^2\rangle$ since we consider $\epsilon(a+b\rightarrow i+j)\propto \sqrt{s-4m_{N_1}^2}\propto v$ in the center-of-mass (CM) frame of $a$ and $b$.  

The complete calculation leads to a troublesome numerical integration and loses the illustration of thermal motion effect, and thus instead we consider an approximation, fixing the value of $W_{ab\to ij}(s=4m_{N_1}^2/(1-v^2))\equiv \int|\mathcal{M}_{tree}(a+b\rightarrow i+j)|^{2}\tilde{\delta}d\Pi_{i}d\Pi_{j}$ at $s= 4m_{N_1}^2$, which amounts to neglecting the thermal motion there. This is reasonably good since the correction is at the higher order of $v^2\ll 1$ in the decoupling region of dark matter. Then, $\delta\gamma^{a b}_{i j }$ can be factorized as 
\begin{align}\label{factorize}
    \delta\gamma^{a b}_{i j }\approx \epsilon_T   
    \frac{n_a^{eq}n_b^{eq}}{4m_{N_1}^2}W_{ab\to ij}(s=4m_{N_1}^2)\equiv \epsilon_T   
    \frac{n_a^{eq}n_b^{eq}}{4m_{N_1}^2}W_0,
\end{align}
with the thermally averaged CP violation parameter defined as
\begin{align}\label{reactionrate1}
   \epsilon_T\equiv \langle\epsilon(s)\rangle =  \frac{4m_{N_1}^2}{n_a^{eq}n_b^{eq}}\int d\Pi_{a}f^{eq}_{a}d\Pi_{b}f^{eq}_{b} \epsilon(a+b\rightarrow i+j).
\end{align}
Note that it is not Lorentz invariant due to the density factor. 

Actually, in BEs, we just want a generalized Eq.~(\ref{crossCPV}) beyond the static limit, namely $\langle \sigma_{ab\to ij} v\rangle-\langle \sigma_{\bar a\bar b\to \bar i\bar j} v\rangle\propto \epsilon_T$, and it can be approximately established as the following. Note that $\epsilon(s)$ can also be written as
\begin{align}
\epsilon(s)=\frac{\sigma_{ab\to ij}v-\sigma_{\bar a\bar b\to \bar i\bar j}v}{\sigma_{ab\to ij}v+\sigma_{\bar a\bar b\to \bar i\bar j}v}.
\end{align}
It is equivalent to the one defined at the amplitude level in Eq.~(\ref{CPV:amp}). Now, let us take the thermal average as in Eq.~(\ref{reactionrate1}) for both sides, to get
\begin{align}
\epsilon_T &=\left\langle\frac{\sigma_{ab\to ij}v-\sigma_{\bar a\bar b\to \bar i\bar j}v}{\sigma_{ab\to ij}v+\sigma_{\bar a\bar b\to \bar i\bar j}v} \right \rangle \approx\frac{\left\langle\sigma_{ab\to ij}v\right \rangle-\left\langle\sigma_{\bar a\bar b\to \bar i\bar j}v\right \rangle}{\left\langle\sigma_{ab\to ij}v\right \rangle+\left\langle\sigma_{\bar a\bar b\to \bar i\bar j}v\right \rangle}.
\end{align}
Therefore, we have a similarity to Eq.~(\ref{crossCPV}),
\begin{align}\label{CPV:BE}
 \left \langle \sigma_{\overline{N_1}\overline{N_1}\rightarrow S_1h} v\right \rangle=\left\langle \sigma_{S_1^*h\rightarrow N_1 N_1} v\right \rangle\equiv\left \langle \sigma_b v\right \rangle  \approx \frac{1-\left \langle \epsilon(T)\right \rangle}{1+\left \langle \epsilon(T)\right \rangle}\left \langle\sigma_{N_1N_1\rightarrow S_1^*h} v\right \rangle
= 
 (1-2\left \langle \epsilon(T)\right \rangle)\left \langle \sigma_a v\right \rangle,
\end{align}
where the first equality is due to the CPT invariance. 

 \subsubsection{Can we have a sufficiently large thermally average CP-violation parameter ?}\label{CPV:analysis}

To gain a numerical impression on the leading effect of the thermal particle motion on CP violation, let us analyze the numerical result of $\epsilon_T$, which in the realistic model receives several contributions. It is illustrative to start from $\epsilon_{12}$ in the limit $\lambda_{sh12}\to 0$ in Eq.~(\ref{CPVs}): 
\begin{align}
    \epsilon_{12}(s)\to -\frac{Im[\lambda_0^*\lambda_1]}{|\lambda_0|^2}\tilde\epsilon_{12}(s)\quad {\rm with} ~~\tilde\epsilon_{12}(s)= \frac{1}{\pi}\frac{\sqrt{(\frac{s}{4}-m_{N_{1}}^2)s}(\frac{s}{2}-m_{N_{1}}^2)}{(s-m_{S_2}^2)m_{N_{1}}^2}.
\end{align}
The numerator of $\tilde\epsilon_{12}(s)$ suffers the static suppression as explained before.  In this limit, the part of pure couplings is estimated to be 
\begin{align}
    \frac{{\rm Im}[\lambda_0^*\lambda_1]}{|\lambda_0|^2}=\frac{{\rm Im}[\lambda_{R1}^*\lambda_{sh1}\lambda_{R2}\lambda_{R1}\lambda_{R2}^*\lambda_{sh1}]}{|\lambda_{R1}\lambda_{sh1}^*|^2}=0.
\end{align}
It is seen that, in the magnitude, the tree level couplings are cancelled by the loop couplings, and thus ${{\rm Im}[\lambda_0^*\lambda_1]}/{|\lambda_0|^2}$ does not have an opportunity to be enhanced. 

However, the situation becomes different if one considers other interferences. For instance, we move to the more interesting resonant scenario for heavy dark matter and consider $\epsilon_{32}(s)$ which in the resonant region approximately $\propto\tilde \epsilon_{32}(s)$, whose expression is obtained by replacing $m_{S_2}$ with $m_{S_1}$ in $\tilde\epsilon_{12}(s)$. While the pure couplings of  $\epsilon_{32}(s)$ now takes the form of  
\begin{align}
    \frac{{\rm Im}[\lambda_0^*\lambda_1]}{|\lambda_0|^2}=\frac{{\rm Im}[\lambda_{R2}^*\lambda_{sh12}\lambda_{R2}\lambda_{R2}\lambda_{R1}^*\lambda_{sh1}^*]}{|\lambda_{R1}\lambda_{sh12}^*|^2}\to \frac{|\lambda_{R1}||\lambda_{R2}||\lambda_{sh1}|}{|\lambda_{sh12}|} e^{i\beta},
\end{align}
where the magnitudes of tree-level couplings are not canceled and thus resonant enhancement may work by taking $\lambda_{sh12}\ll 1$. 


We now focus on $\tilde\epsilon(s)$. It can be simplified into a one-dimension integration over $s$ in the CM frame. As $T$ is close to the dark matter decoupling temperature, the dark matter particles approximately follow the classical Maxwell distribution  $f(E_{N_1})=1/\exp(E_{N_1}/T)$ where the temperature $T$ equals to that of the plasma Ref.~\cite{Davidson:2008bu}. Following the method in Ref.~\cite{Davidson:2008bu}, we isolate the thermal motion of the center of mass for two colliding dark matter particles $a+b\to i+j$ by inserting the identity $\begin{aligned}1=\int d^4Q\delta^4(Q-p_a-p_b)\end{aligned}$ into Eq.~(\ref{reactionrate1}), with $Q_0=E_{a}+E_{b}$. Further, the two-body phase space can be manipulated via 
\begin{equation} 
\int\tilde{\delta}d\Pi_pd\Pi_q=\int\frac{|\vec{p}_p|}{16\pi^2\sqrt{s}}d\Omega_p=\frac{|\vec{p_p}-\vec{p_q}|}{8\pi\sqrt{s}}=\frac{\sqrt{(p_p\cdot p_q)^2-m_p^2m_q^2}}{4\pi s}.
\end{equation}
Then, we have
\begin{align}
    \left\langle \tilde\epsilon_{32} (T)\right\rangle =\frac{m_{N_1}^2}{n_a^{eq}n_b^{eq}}\int\frac{dQ_{0}d^{3}Q}{(2\pi)^{4}}\frac{e^{-Q_{0}/T}}{\pi s}\sqrt{s(\frac{s}{4}-m_{N1}^2)}\epsilon_{32}(a+b\rightarrow i+j),
\end{align}
Using $d^3Q=\sqrt{Q_0^2-s}dsd\Omega/2$, it becomes
\begin{align}
    \left\langle \tilde\epsilon_{32}(T)\right\rangle 
&=\frac{m_{N_1}^2}{n_a^{eq}n_b^{eq}}\frac{1}{32\pi^5}\int \frac{4}{s} dsd\Omega\int_{\sqrt{s}}dQ_0e^{-Q_0/T}\sqrt{Q_0^2-s}\sqrt{s(\frac{s}{4}-m_{N1}^2)}\epsilon_{12}(a+b\rightarrow i+j)  
\\ \nonumber
&= \frac{m_{N_1}^2}{n_a^{eq}n_b^{eq}}  \frac{T}{8\pi^4}\int_{4m_{N_1}^2} dsK_1\left(\frac{\sqrt{s}}{T}\right)\sqrt{\frac{s}{4}-m_{N1}^2}\epsilon_{32}(s).
\end{align} 
Further introducing the dimensionless couplings $x\equiv \sqrt{s}/T$, $y\equiv m_{S_1}/T$ and $2y<z\equiv m_{N_1}/T<y$ and utilizing the Bessel function: $zK_1(z)\to \sqrt{\frac{\pi z}{2}}e^{-z}$ and $z^2K_2(z)\to (15/8+z)\sqrt{\frac{\pi z}{2}}e^{-z}$  for $z\gg1$, we arrive
\begin{equation} 
  \left\langle \tilde\epsilon_{32}(T)\right\rangle 
\approx \frac{ e^{2z}}{8g_N^2(15/8+z)^2z\pi\sqrt{\frac{\pi }{2}}}\int_{2z} \sqrt{x}e^{-x}\sqrt{x^2-4z^2}\frac{x\sqrt{(x^2-4z^2)}(x^2-2z^2)}{(x^2-y^2)} dx,
 \end{equation}
with $g_N=2$ for the fermionic dark matter excluding its antiparticle. 

\begin{figure}[htbp]
\centering
\includegraphics[width=0.6\textwidth]{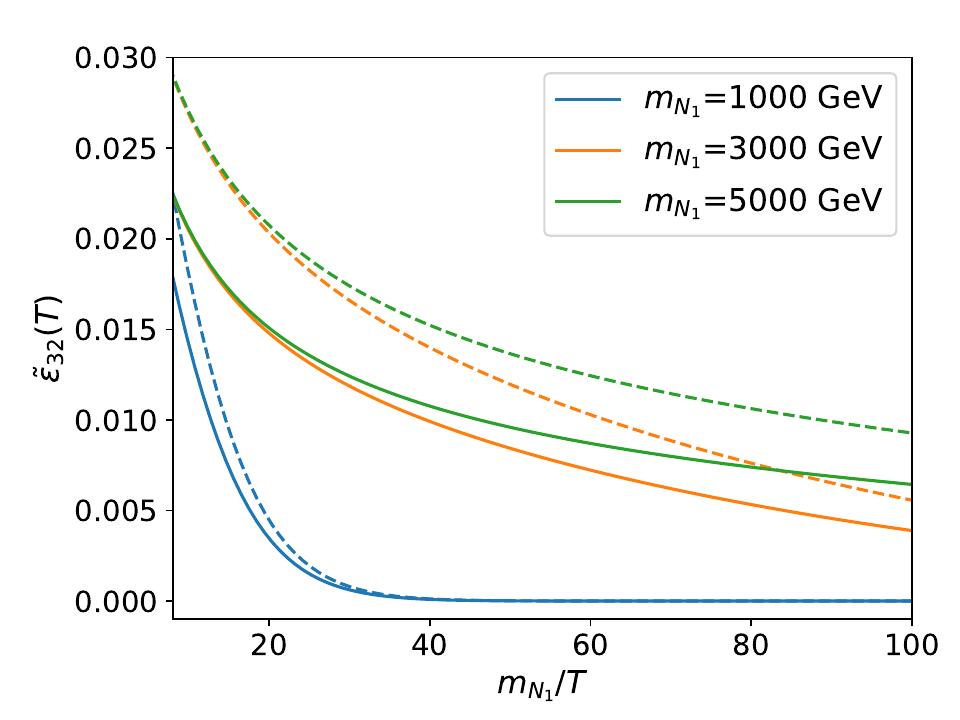}
\caption{The thermally averaged CP-violation parameter $\tilde\epsilon_{32}(T)$ as a function of $m_{N_1}/T$ near the dark matter decoupling region. In the plot, we choose several dark matter mass benchmark values and take two ratios $m_{S_1}/m_{N}=$1.2 (solid line) and 1.5 (dotted line). }
\label{cp-vio}
\end{figure}

The numerical samples are demonstrated in Fig.~\ref{cp-vio}.  From the figure, we can see that $\tilde\epsilon_{32}(T)$ increases correspondingly with the mass of dark matter or the ratio ${m_{S_1}}/{m_{N_1}}$. The magnitude of $\tilde\epsilon_{32}(T)\sim {\cal O}(10^{-2})$ has a moderate suppression as expected, which requires an accordingly enhancement from the coupling part to arrive $\epsilon_{32}(T)\sim {\cal O}(0.1)$. According to the analysis above, this enhancement is feasible in the resonant region.

\section{ A preliminary numerical study}

In this section, we solve the coupled BEs Eq.~(\ref{BEs:compact1}-\ref{BEs:compact2}) numerically and display the results of the model. Our task is to explore the parameter space that can correctly reproduce the dark matter relic density $\Omega_{\rm DM} h^2\approx 0.11$ and the baryonic asymmetry as well. To verify our code for solving BEs, we compare the dark matter relic density calculated using our python code with that calculated using the popular package micrOMEGAs~\cite{Belanger:2020gnr}. We find that the discrepancy is below $2\%$ when the CP-violation parameter is set to zero, which is acceptable.

There are several free parameters crucial to our discussions: the masses of $N_{1}$, $S_1$ and the coupling parameters ($\lambda_{sh1}$, $\lambda_{R1}$, $y_{1}$). In determining the feasible parameter space where the correct relic density and asymmetry can be achieved, we perform scans over two parameters at a time while holding others fixed.

\begin{figure}[htbp]
\centering
\includegraphics[width=0.48\textwidth]{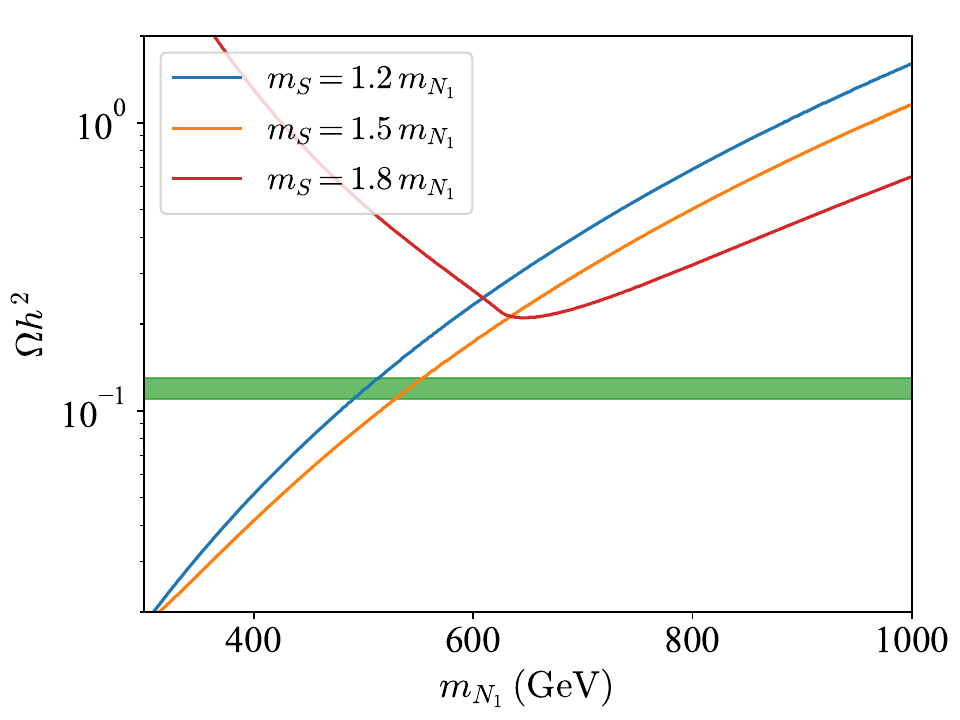}
\includegraphics[width=0.48\textwidth]{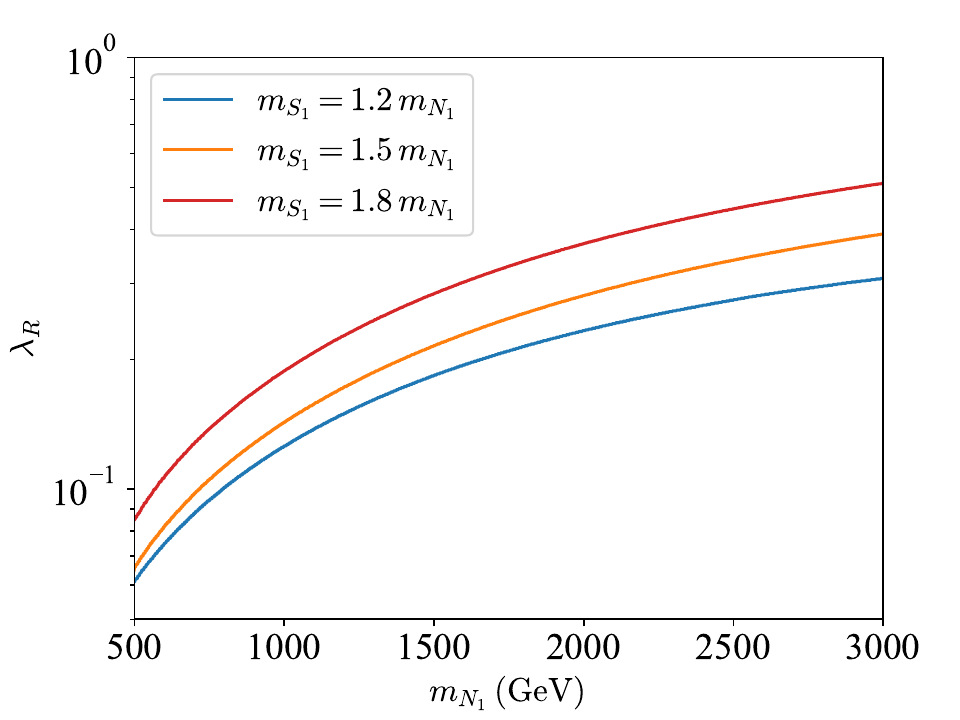}
\caption{Left panel: The DM relic density $\Omega h^2$ and Dark matter mass for the different ratios of mediator mass to dark matter mass with $\lambda_R=1$ fixed. The green band marks the observed value~\cite{arbey2021dark}. Right panel: Dark matter mass vs. $\lambda_R$ under resonant enhancement with $m_{S_2}\approx 2m_{N_{1}}$.}
\label{relic}
\end{figure}

We first figure out the preferred parameter space for correct dark matter relic density, which is sensitive to all the parameters except $y_1$. Actually, without turning on large CP violation, the light dark matter region (below 1TeV) has been detailedly studied in our previous work~\cite{Guo:2021rre} using micrOMEGAs. Co-annhilation effect plays an important role, but we here exclude it by setting $m_{S_1}>1.2 m_{N_1}$ to avoid the subtle impact on dark matter asymmetry~\footnote{In our current consideration, dark matter does not have DM and anti-DM annihilation, otherwise it will brings a substantial difference to dark matter relic density and as well the indirect detection signature. However, it is an important and interesting scenario worthy of further study elsewhere.}.  Thus, we only show the typical behaviors of relic density varying with dark matter mass in the left panel of  Fig.~\ref{relic}, fixing other parameters. Note that for $m_{S_1}=1.8 m_{N_1}$,  the curve of DM relic density curves upwards as DM decreases within the range of DM mass below 500 GeV. The reason is nothing but that the mass relation tends to violate the condition Eq.~(\ref{window}):   $m_{S_1}/m_{N_1}+m_h/m_{N_1}=1.8+m_h/m_{N_1}<2$, thus squeezing the available phase space for DM annihilation.

The heavy dark matter of multi-TeV is of more interest since it is related to baryogenesis. To leave a window to transfer the lepton asymmetry to baryons, the DM freeze-out temperature is required to be higher than around the sphaleron freeze-out temperature $T_{\text{FO}}\sim 100$ GeV, which in turn sets a lower bound on the dark matter mass, $m_{N_{1}}\gtrsim x_{f,N_1} T_{\text{FO}}$ with $x_{f,N_1}\approx 30$ larger than the usual typical value explained before. In general, to make a multi-TeV scale fermionic dark matter obtain the correct relic density, the required couplings will become unreasonably large. Therefore, we turn to the more or less fine tuned case that dark matter annihilation meets a resonance, and the only candidate is the heavier companion $S_2$. With resonant enhancement, the heavy dark matter is feasible with a relatively small $\lambda_{R1}$, shown in the right panel of Fig.~\ref{relic}. Unfortunately, in this scenario, the spectra of the scalar sector are pushed into the multi-TeV region and difficult to probe at colliders.


 

To show the range of the allowed couplings, we select a set of representative values for the particle masses and CP-violation parameters in various cases. That means we obtain the appropriate range of couplings by treating relic density as a function of all masses and all couplings. We plot the values of the coupling parameters in Fig.~\ref{relic} below. From the figure, we can see that as the mass of dark matter increases, $\lambda_{R1}$ increases accordingly.

\begin{figure}[htbp]
\begin{center}
\includegraphics[width=0.5\textwidth]{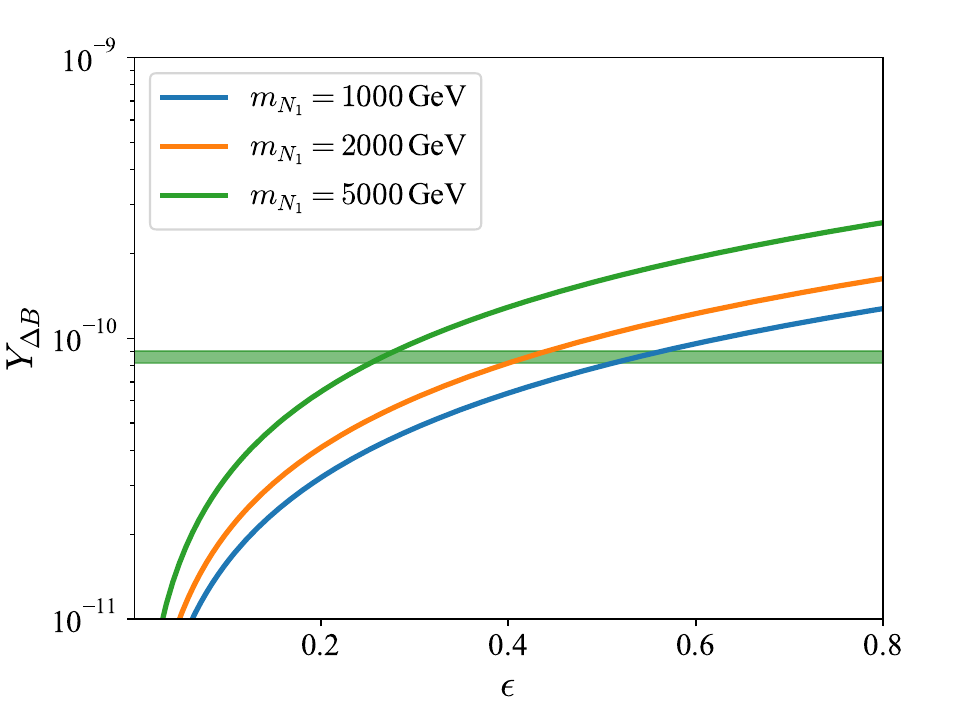}\includegraphics[width=0.5\textwidth]{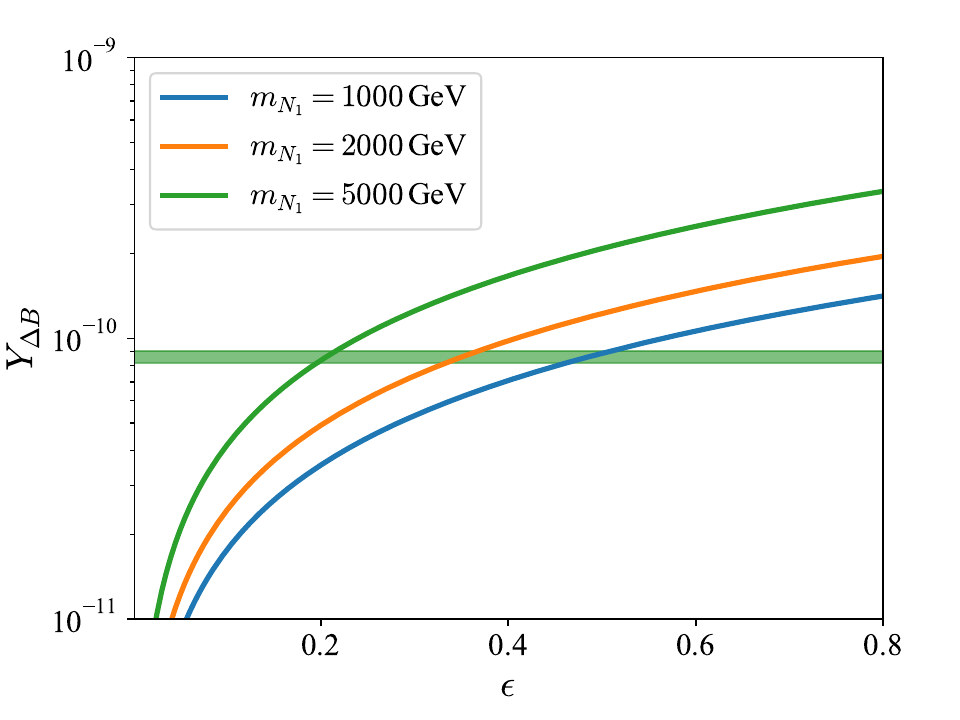}
\end{center}
\caption{Plots of the abundance of baryon asymmetry with dark matter mass and parameters with horizontal bands imply an estimate of baryon asymmetry~\cite{Planck:2013pxb}, where $\lambda_{sh1}v$=1 GeV, $\lambda_{R1}=1$, $y_{1}=1$, ${m_{S_1}}/{m_{N_1}}=1.2$ (left) and ${m_{S_1}}/{m_{N_1}}=1.5$ (right).}
\label{abc}
\end{figure}

When it comes to baryon asymmetry, we find that a suitable baryon asymmetry density can be obtained within a certain dark matter mass range which is shown in Fig.~\ref{abc}. We fixed the size of the coupling parameter except $m_{N_1}$ and the CP-violation parameter. From the figure, it is seen that if we want to get the right baryon asymmetry, the required cp-violation parameter will become smaller as the mass of dark matter increases.

\section{ Conclusion and Discussion}

 The WIMP has been a leading candidate for DM model-building. However, the typical WIMP DM suffers the strong exclusion from direct detection experiments. The $Z_{N\geq 3}$ symmetric DM-companion models provide us with safe WIMP DM candidates, and in this work, we find that they also naturally seed DM asymmetry and thus may answer to the BAU puzzle after introducing a proper link into SM.

In this article, we consider the benchmark $Z_3$ DM-companion model as a concrete example, where DM is a Dirac fermion and its companion is a complex scalar, servicing as the portal to SM. Moreover, we add a leptonic portal coupling to DM with the help of a scalar doublet. Such a simple extension gives all the ingredients to solve the BAU puzzle, which includes generating CP asymmetry by DM annihilation processes, CP asymmetry processes freezing out as DM freeze-out and transferring asymmetry of Dark sector to neutrinos by DM companion decaying. We write out the BEs of various particle number densities, evaluate the CP-violation parameters and analyze the favored parameter space of our model. For the interesting multi-TeV scale DM, the model with $\epsilon$ around ${\cal O}(0.1)$ works. In addition, it is worth noting that the CP-violation parameter will be suppressed highly if the thermal motion of the initial particle is not taken into account. When considering the thermal average parameter $\epsilon_{32}(T)$, a relatively small $\lambda_{sh12}$ is required to obtain reasonable results. This coupling can be very small due to DM annihilation with resonant enhancement.



There are still some open questions here. First, the DM-companion model allows for co-annihilation, which has not been considered in our article for the time being. This is subject to further study and important effect on DM asymmetry is expected. Second, our model actually produces Majorana neutrino masses at two-loop level, and we have considered a sufficiently small lepton portal coupling to make them sufficiently light. However, it is of great importance to investigate if the neutrino phenomenology can be accommodated in a unified framework with matter asymmetry and safe WIMP DM, which is capable of avoiding the stringent constraints from DM direct detection. 



\acknowledgments
This work is supported in part by the National Natural Science Foundation of China (12175082,11775093) and the Fundamental Research Funds for the Central Universities (CCNU22LJ004).

\newpage

\appendix 

\section{The evaluation of CP-violataion}\label{CPV}

The usual CP-violation parameter at zero temperature is defined at the level of amplitude, receiving the tree and one-loop contributions, whose matrix elements can be factorized into a coupling constant part $\lambda$ and an amplitude part $I$ as the following,
\begin{align}
    M=M_{0}+M_{1}=\lambda_{0}I_{0}+\lambda_{1}I_{1}.
\end{align}
Then, the CP-violation parameter can be expressed as: 
 \begin{align}\label{cpvp}
     \epsilon\equiv\frac{|M|^2_{N_{1}N_{1}\rightarrow S_1^*h}-|M|^2_{\bar{N}_{1}\bar{N}_{1}\rightarrow S_1h}}{|M|^2_{N_{1}N_{1}\rightarrow S_1^*h}+|M|^2_{\bar{N}_{1}\bar{N}_{1}\rightarrow S_1h}}
    \approx-2\frac{{\rm Im}[\lambda_0^*\lambda_1]}{|\lambda_0|^2}
\frac{{\rm Im}[I_0^*I_1]}{|I_0|^2}.
      \end{align}
By applying the Cutkosky rules, one can calculate the imaginary part of the amplitude ${\rm Im}[I_0^*I_1]$, arising due to the on shell of loop particles, ${\rm Im}{I_0^*I_1}=\frac{1}{2}I_0^*\sum_{cuttings}I_1=I_0^*{\rm Im}{I_1}$.

Let us apply the above procedure to the scattering $N_1(p_A)N_1(p_B)\rightarrow S_1^*h$, where the intermediate particles running in the loop are the same with the initial particles. We take the contribution specified in Fig.~\ref{tree-loop8} as an example, then
 \begin{align}\label{}
     {\rm Im}[\lambda_0^*\lambda_1]={\rm Im}[\lambda_{R1}^*\lambda_{sh1}\lambda_{R2}\lambda_{R1}\lambda_{R2}^*\lambda_{sh1}^*]v^2.
      \end{align}
In the center of mass frame, the four-momentum for the initial and intermediate particles read
\begin{align}
    p_A =(E, \vec{p}),\quad p_B =(E, -\vec{p}),\quad  p_1 =(E, \vec{k}),\quad p_2 =(E, -\vec{k}),
\end{align}
so $s=(p_1+p_2)^2=4 E^2=4m_{N_1}^2+4|\vec p|^2$. Then, the tree-level amplitude reads
\begin{align}
    I_0^*=\frac{\overline{v(p_A)}(1-\gamma^5)u(p_B)}{q^2-m_{S_1}^2}.
\end{align}
Using the cutting rule, we have 
\begin{align}
   \frac{1}{2}\sum_{\text{cuttings}}I_1&=
    \frac{1}{2}\int d\Pi_f I_0(N_{1}N_{1}\rightarrow N_{1}N_{1})I_0(N_{1}N_{1}\rightarrow S_1^*h),
    \\\nonumber
    &=\frac{1}{2}\int d\Pi_f \frac{\overline{u(p_B)}(1+\gamma^5)v(p_A)\overline{v(p_1)}(1-\gamma^5)u(p_2)}{q^2-m_{S_2}^2}\times \frac{\overline{u(p_2)}(1+\gamma^5)v(p_1)}{q^2-m_{S_1}^2},
\end{align}
where the momentum of the intermediate particle with corresponding spin is $p_1$ and $p_2$, Therefore, we get the expression for this term in the Eq.~(\ref{cpvp}) 
\begin{equation}
\begin{split}\label{nn}
   {\rm Im}{I_0^*I_1}&= \frac{\overline{v(p_A)}(1-\gamma^5)u(p_B)}{q^2-m_{S_1}^2}\\
   &\frac{1}{2}\times \int d\Pi_f \frac{\overline{u(p_B)}(1+\gamma^5)v(p_A)\overline{v(p_1)}(1-\gamma^5)u(p_2)}{q^2-m_{S_2}^2} \frac{\overline{u(p_2)}(1+\gamma^5)v(p_1)}{q^2-m_{S_1}^2},\\
    &=\frac{1}{2}\frac{v^28p_A\cdot p_B}{(q^2-m_{S_1}^2)^2(q^2-m_{S_2}^2)}\int d\Pi_f (8p_1\cdot p_2)=\frac{1}{\pi}\frac{\sqrt{s(s-4m_{N1}^2)}(s-2m_{N1}^2)}{(s-m_{S_1}^2)^2(s-m_{S_2}^2)}.
\end{split}
\end{equation}
which depends on the initial momentum of the dark matter pair. In particular, the size has a nonrelativistic suppression.

\section{Vanishing CP-violation in the electroweak channel}


Let us move to the other channel with $h$ replaced by the $Z$ boson. The relevant Lagrangian reads
\begin{align}
    {\cal L}\supset &-Z^\mu(S\partial_\mu S ^*+S^*\partial_\mu S)
    \end{align}
Then, we calculate another interface process of $N(p_A)_1N(p_B)_1\rightarrow S_1(p_4)^*Z(p_5)$ and its loop corrections as follows,
\begin{align}
{\rm Im}{I_0^*I_1}=\frac{1}{2}I_0^*\sum_{\text{cutting}s}I_1=I_0^*{\rm Im}{I_1},
\end{align}
where $I_0^*=\frac{\overline{u(p_A)}(1-\gamma^5)u(p_B)}{q^2-m_{S_1}^2} (p_3-p_4)^\mu \epsilon ^*_\mu (p_5,\lambda).$ For example, we deal with a process $N(p_A)_1N(p_B)_1\rightarrow N(p_1)_1N(p_2)_1 \rightarrow S_1^*Z$ which means dark matter candidate is the intermediate particle in the loop. Thus,
\begin{equation}\label{zbc}
\begin{split}
    {\rm Im}{I_1}&\ni
    \frac{1}{2}\int d\pi_f I_0(N_{1}N_{1}\rightarrow N_{1}N_{1})I_0(N_{1}N_{1}\rightarrow S_1^*Z),\\
    &=\frac{1}{2}\int d\pi_f \frac{\overline{u(p_B)}(1+\gamma^5)u(p_A)\overline{u(p_1)}(1-\gamma^5)u(p_2)}{q^2-m_{S_2}^2}\\
    &\times \frac{\overline{u(p_2)}(1+\gamma^5)u(p_1)}{q^2-m_{N_{1}}^2}(p_3-p_4)^\mu \epsilon_\mu (p_5,\lambda).
\end{split}
\end{equation}
Using the Eq.~\ref{zbc}, we can get a representation of  ${\rm Im}{I_0^*I_1}$ as follows

\begin{equation}\begin{aligned}
\mathrm{Im}I_{0}^{*}I_{1}& \ni\frac{u(p_A)(1-\gamma^5)u(p_B)}{q^2-m_{S_1}^2}(p_3-p_4)^\mu\epsilon_\mu^*(p_5,\lambda)  \\
&\frac{1}{2}\times\int d\pi_{f}\frac{\overline{{u(p_{B})}}(1+\gamma^{5})u(p_{A})\overline{{u(p_{1})}}(1-\gamma^{5})u(p_{2})}{q^{2}-m_{S_{2}}^{2}} \\
&\times\frac{\overline{{u(p_{2})}}(1+\gamma^{5})u(p_{1})}{q^{2}-m_{S_{1}}^{2}}(p_{3}-p_{4})^{\mu}\epsilon_{\mu}(p_{5},\lambda), \\
&=\frac{1}{2}\frac{8p_{A}\cdot p_{B}}{(q^{2}-m_{S_{1}}^{2})^{2}(q^{2}-m_{S_{2}}^{2})}(p_{3}-p_{4})^{\mu}\epsilon_{\mu}\epsilon_{\nu}(p_{3}-p_{4})^{\nu}\int d\pi_{f}(8p_{1}\cdot p_{2}).
\end{aligned}\end{equation}

In the frame of center mass, $p_A =(E, \vec{p})$, $p_B =(E, -\vec{p}), p_1 =(E, \vec{k})$, $p_2 =(E, -\vec{k})$, $p_3 =(2E, 0)$, $p_4 =(E_1, \vec{t})$, $p_5 =(E_2, -\vec{t})$,and then $s=(p_1+p_2)^2=4 E^2$. Therefore,
\begin{align}\label{zz}
     {\rm Im}{I_0^*I_1}&\ni\frac{i}{\pi}\frac{(m_z^2-m_z^2)\sqrt{s(s-4m_{N1}^2)}(s-2m_{N1}^2)}{(s-m_{S_1}^2)^2(s-m_{S_2}^2)}=0.
\end{align}
According to the above, we find that the CP-violation parameter will vanish for either Eq.~(\ref{nn}) or Eq.~(\ref{zz}). 

\section{The detailing of Boltzmann Equations}

    

In this appendix, using Eq.~(\ref{dif}) and Eq.~(\ref{rate}), we expand the compact BEs given in Eq.~(\ref{BEs:compact1}-\ref{BEs:compact2}) to present the details of collision terms :
\begin{align}\label{BEs:expand1}
\frac{\mathrm{d}Y_{S_1^*}}{\mathrm{d}z_1}=&\frac{m_1^3}{z_1^2 H(m_1)}[Y_{N_{1}} Y_{N_{1}}\left \langle\sigma_b v\right \rangle-Y_{S^*_1}Y_{h}\left \langle\sigma_a v\right \rangle+Y_{\nu_1} e^{-\frac{z_1 \Delta m }{m_1}}\frac{z_1^3}{m_1^3}\Gamma_{D_1}-Y_{S_1^*}\frac{z_1^3}{m_1^3}\Gamma_{D_1}\\
    \nonumber
    &+Y_{\bar{N}_{1}}Y_{\nu_1}\left \langle\sigma_c v\right \rangle-Y_{S_1^*}Y_{h}\left \langle\sigma_c v\right \rangle],\\
    \frac{\mathrm{d}Y_{S_1}}{\mathrm{d}z_1}=&\frac{m_1^3}{z_1^2 H(m_1)}[Y_{\bar{N}_{1}} Y_{\bar{N}_{1}}\left \langle\sigma_a v\right \rangle-Y_{S_1}Y_{h}\left \langle\sigma_b v\right \rangle+Y_{\overline{\nu_1}} e^{-\frac{z_1 \Delta m }{m_1}}\frac{z_1^3}{m_1^3}\Gamma_{D_1}-Y_{S_1}\frac{z_1^3}{m_1^3}\Gamma_{D_1}\\
    \nonumber
    &+Y_{N_{1}}Y_{\overline{\nu_1}}\left \langle\sigma_c v\right \rangle-Y_{S_1}Y_{h}\left \langle\sigma_c v\right \rangle],\\
    \frac{\mathrm{d}Y_{S_2^*}}{\mathrm{d}z_2}=&\frac{m_2^3}{z_2^2 H(m_2)}[Y_{N_{1}} Y_{N_{1}}\left \langle\sigma_b v\right \rangle-Y_{S_2^*}Y_{h}\left \langle\sigma_a v\right \rangle+Y_{\nu_1} e^{-\frac{z_2 \Delta m }{m_2}}\frac{z_2^3}{m_2^3}\Gamma_{D_2}-Y_{S_2^*}\frac{z_2^3}{m_2^3}\Gamma_{D_2}\\
    \nonumber
    &+Y_{\bar{N}_{1}}Y_{\nu_1}\left \langle\sigma_c v\right \rangle-Y_{S_2^*}Y_{h}\left \langle\sigma_c v\right \rangle],\\
    \frac{\mathrm{d}Y_{S_2}}{\mathrm{d}z_2}=&\frac{m_2^3}{z_2^2 H(m_2)}[Y_{\bar{N}_{1}} Y_{\bar{N}_{1}}\left \langle\sigma_a v\right \rangle-Y_{S_2}Y_{h}\left \langle\sigma_b v\right \rangle+Y_{\overline{\nu_1}} e^{-\frac{z_2 \Delta m }{m_2}}\frac{z_2^3}{m_2^3}\Gamma_{D_2}-Y_{S_2}\frac{z_2^3}{m_2^3}\Gamma_{D_2}\\
    \nonumber
    &+Y_{N_{1}}Y_{\overline{\nu_1}}\left \langle\sigma_c v\right \rangle-Y_{S_2}Y_{h}\left \langle\sigma_c v\right \rangle],\\
    \frac{\mathrm{d}Y_{N_{1}}}{\mathrm{d}z_N}=&\frac{m_{N_{1}}^3}{z_N^2 H(m_{N_{1}})}[2Y_{S_1^*}Y_{h}\left \langle\sigma_a v\right \rangle-2Y_{N_{1}} Y_{N_{1}}\left \langle\sigma_b v\right \rangle+Y_{S_1}\frac{z_N^3}{m_{N_{1}}^3}\Gamma_{D_1}-Y_{\nu_1} e^{-\frac{z_N \Delta m }{m_{N_{1}}}}\frac{z_N^3}{m_{N_{1}}^3}\Gamma_{D_1} \nonumber \\
    &+Y_{S_1}Y_{h}\left \langle\sigma_c v\right \rangle-Y_{N_{1}}Y_{\overline{\nu_1}}\left \langle\sigma_c v\right \rangle+2Y_{S_2^*}Y_{h}\left \langle\sigma_a v\right \rangle-2Y_{N_{1}} Y_{N_{1}}\left \langle\sigma_b v\right \rangle \nonumber \\
    &+Y_{S_2}\frac{z_N^3}{m_{N_{1}}^3}\Gamma_{D_2}-Y_{\nu_1} e^{-\frac{z_N \Delta m }{m_1}}\frac{z_N^3}{m_{N_{1}}^3}\Gamma_{D_2}+Y_{S_2}Y_{h}\left \langle\sigma_c v\right \rangle-Y_{N_{1}}Y_{\overline{\nu_1}}\left \langle\sigma_c v\right \rangle],\\
    \frac{\mathrm{d}Y_{\bar{N}_{1}}}{\mathrm{d}z_N}=&\frac{m_{N_{1}}^3}{z_N^2 H(m_{N_{1}})}[2Y_{S_1}Y_{h}\left \langle\sigma_b v\right \rangle-2Y_{\bar{N}_{1}} Y_{\bar{N}_{1}}\left \langle\sigma_a v\right \rangle+Y_{S_1^*}\frac{z_N^3}{m_{N_{1}}^3}\Gamma_{D_1}-Y_{\overline{\nu_1}} e^{-\frac{z_N \Delta m }{m_{N_{1}}}}\frac{z_N^3}{m_{N_{1}}^3}\Gamma_{D_1}\nonumber \\
    &+Y_{S_1^*}Y_{h}\left \langle\sigma_c v\right \rangle-Y_{\bar{N}_{1}}Y_{\nu_1}\left \langle\sigma_c v\right \rangle+2Y_{S_2}Y_{h}\left \langle\sigma_b v\right \rangle-2Y_{\bar{N}_{1}} Y_{\bar{N}_{1}}\left \langle\sigma_a v\right \rangle \nonumber \\
    &+Y_{S_2^*}\frac{z_N^3}{m_{N_{1}}^3}\Gamma_{D_2}-Y_{\overline{\nu_1}} e^{-\frac{z_N \Delta m }{m_{N_{1}}}}\frac{z_N^3}{m_{N_{1}}^3}\Gamma_{D_2}+Y_{S_2^*}Y_{h}\left \langle\sigma_c v\right \rangle-Y_{\bar{N}_{1}}Y_{\nu_1}  \langle\sigma_c v \rangle],
    \label{BEs:expand2}
 \end{align}
 where $\sigma_a$, $\sigma_b$, $\sigma_c$ and $\sigma_d$ orderly denote the cross section of the processes  $\overline{N_1}\overline{ N_1}\rightarrow S_1h$, $S_1h\rightarrow \overline{N_1}\overline{ N_1}$, $S_1^*h\rightarrow \overline{N_1 }\nu_1$ and $S_1^*\rightarrow \overline{N_1 }\nu_1$. According to the definition of $\epsilon$, one has
\begin{align}
\epsilon=\frac{\sigma_{N_{1}N_{1}\rightarrow S_1^*h}v-\sigma_{\bar{N}_{1}\bar{N}_{1}\rightarrow S_1h}v}{\sigma_{N_{1}N_{1}\rightarrow S_1^*h}v+\sigma_{\bar{N}_{1}\bar{N}_{1}\rightarrow S_1h}v}.
\end{align}

Besides, we evaluate the expression for $\sigma_{N_1N_1\rightarrow S_1^*h}v$ and $\Gamma_{D2}$

\begin{align}
\sigma_{N_1N_1\rightarrow S_1^*h}v&= \frac{|\vec{p}_h|\lambda_{R1}^2\lambda_{sh1}^2v^2m_{N_1}^2}{8\pi\sqrt{s}^3(s-m_{s_1}^2)^2},\\
\nonumber
\Gamma_{D2}&=\frac{1}{16\pi}\frac{y_1^2(m_{S1}^2-m_{N}^2)^2}{m_{S1}^3}
\end{align}
where $|\vec{p}_h|$ is momentum of Higgs.

If we take the effect of the interaction $S_1S_1^*\rightarrow hh$ into consideration, the first line of the above equations can be rewritten as

\begin{align}
\frac{\mathrm{d}Y_{S_1^*}}{\mathrm{d}z_1}=&\frac{m_1^3}{z_1^2 H(m_1)}[Y_{N_{1}} Y_{N_{1}}\left \langle\sigma_b v\right \rangle-(Y_{N_1}^{eq})^2\frac{Y_{S^*_1}}{Y_{S^*_1}^{eq}}\left \langle\sigma_a v\right \rangle+Y_{\nu_1} e^{-\frac{z_1 \Delta m }{m_1}}\frac{z_1^3}{m_1^3}\Gamma_{D}-Y_{S_1}\frac{z_1^3}{m_1^3}\Gamma_{D} \nonumber \\
    &+Y_{N_{1}}Y_{\overline{\nu_1}}\left \langle\sigma_c v\right \rangle-Y_{S_1}Y_{h}\left \langle\sigma_c v\right \rangle+Y_{S_{1}}^{*eq} Y_{S_{1}}^{*eq}\left \langle\sigma_d v\right \rangle-Y_{S_{1}}^*Y_{S_{1}}^*\left \langle\sigma_d v\right \rangle],
 \end{align}
 where $\sigma_d$ denote the cross section of the processes  $S_1S_1^*\rightarrow hh$. Similar operations can be performed on other equations in the BEs.





\end{document}